\documentclass[12pt]{elsarticle}

\usepackage[margin=1.0in]{geometry}
\usepackage{graphicx}
\usepackage{amsmath}
\usepackage{amssymb}
\usepackage{tabularx}
\usepackage{caption}
\usepackage{subcaption}
\usepackage{balance}
\usepackage{mathtools}
\usepackage{url}
\usepackage{changes}
\usepackage{color}

\journal{Ad Hoc Networks}

\allowdisplaybreaks

\begin{document}

\begin{frontmatter}

\title{Network Lifetime Maximization in Wireless Mesh Networks for Machine-to-Machine Communication}

\author[LU,PW]{Emma Fitzgerald
\corref{mycorrespondingauthor}}
\ead{emma.fitzgerald@eit.lth.se}
\cortext[mycorrespondingauthor]{Corresponding author}

\author[PW]{Micha\l \ Pi\'{o}ro}
\ead{m.pioro@tele.pw.edu.pl}

\author[PW]{Artur Tomaszewski}
\ead{a.tomaszewski@tele.pw.edu.pl}

\address[LU]{Department of Electrical and Information Technology, Lund University, Lund, Sweden}
\address[PW]{Institute of Telecommunications, Warsaw University of Technology, Warsaw, Poland}

\begin{abstract}
In this paper we present new optimization formulations for maximizing the network lifetime in wireless mesh networks performing data aggregation and dissemination for machine-to-machine communication in the Internet of Things. We focus on heterogeneous networks in which multiple applications co-exist and nodes may take on different roles for different applications. Moreover, we address network reconfiguration as a means to increase the network lifetime, in keeping with the current trend towards software defined networks and network function virtualization. To test our optimization formulations, we conducted a numerical study using randomly-generated mesh networks from 10 to 30 nodes, and showed that the network lifetime can be increased using network reconfiguration by up to 75\% over a single, minimal-energy configuration. Further, our solutions are feasible to implement in practical scenarios: only few configurations are needed, thus requiring little storage for a standalone network, and the synchronization and signalling needed to switch configurations is low relative to each configuration's operating time.
\end{abstract}


\end{frontmatter}

\section{Introduction}\label{sec:introduction}

In an increasingly wireless world, and in particular with the rise of the
Internet of Things, energy efficiency for end devices is a critical component in
enabling new applications. Taking an application-centric view, it is not the
energy consumption of individual nodes in the network that is the most important
consideration, but rather how long the network as a whole can fulfill its
intended purpose, that is, serve the demands of the application(s) running on
it. This time is called the \emph{lifetime} of the network.

To achieve maximum network lifetime, reconfiguration of the network, in
the form of changing routing and/or which tasks are assigned to which nodes, may be
necessary. For example, with a given set of paths taken by the various
application data flows, some nodes may be more heavily loaded than
others and become bottlenecks, needing to transmit often and draining their
batteries more quickly. Once these critical nodes are out of power, the path(s)
on which they lie will fail, causing a disruption in service. However, in many
cases, especially for mesh networks, it is possible to find other feasible paths
consisting only of nodes that still have some power remaining. In fact, to
achieve the longest possible lifetime, reconfiguration of routing of
traffic demands may need to be performed multiple times.

In future, especially as 5G comes into effect, such reconfiguration will become
more feasible. There is currently a trend towards software-defined
networking and network function virtualization, making telecommunications
networks more flexible and reconfigurable. Moreover, it is typical that end
devices will have a connection to cloud or edge servers, possibly through
multiple different gateways, and therefore do not need to themselves have
sufficient computation power to determine the optimal configuration or
reconfiguration. Instead, this may be done in the cloud or the fog, and then
communicated to the end devices.

In our previous work \cite{fitzgerald2018energy}, we studied the problem of
routing in a wireless mesh network together with data aggregation and
dissemination for machine-to-machine communication, optimizing for minimal total
energy usage (which can equivalently be understood as the minimal average power
consumption by the network nodes). In the current paper, we now present 
complementary, novel optimization formulations for maximum network lifetime (i.e., the
time until the network ceases to be fully operational), in which the network is
able to be reconfigured both in terms of routing of traffic streams, and which
nodes are selected to aggregate and/or disseminate (via multicast transmission)
individual sensor measurements. We examine practical implementation issues
and describe how our approach can be deployed in real networks. We also
conducted a numerical study solving our optimization problems for randomly
generated mesh networks with 10 to 30 nodes. Our results show that the network
lifetime can be increased by up to 75\% compared with configuring the network
for minimum total energy usage, and that relatively few (around 10) different
configurations are needed to achieve the maximum lifetime. This means that the
optimal solutions we find here are feasible to implement in practice, as the
overhead for reconfiguration will be low compared to the total network lifetime.

The contributions of this paper are the following.
\begin{enumerate}
    \item We provide novel optimization formulations for maximizing network
    	lifetime that allow for reconfiguration of routing and node tasks.
    \item We develop a general solution approach to these optimization problems,
    	based on column generation. This allows our approach to be applied to
    	arbitrary network tasks.
    \item We examine the specific case of machine-to-machine communication, in
    	which sensor measurements can be aggregated within the network, and must
    	be disseminated via multicast transmission to multiple destinations. We
    	give an appropriate pricing problem formulation for this application.
    \item Our solution approach is feasible to implement in practice, since only
    	few configurations are used for maximal lifetime, and the requirements
	for signalling and synchronization needed to perform reconfiguration are
	low.
    \item We present results from a numerical study investigating the
    	performance of our optimization approach, and showing that it can
    	provide large improvements in network lifetime for the considered
    	application. We compare performance for maximum network lifetime with
    	that for total energy minimization, and discuss the trade-offs between
    	these two approaches.
    \item We provide tight upper and lower bounds for the optimal solutions to
    	our formulations, as well as a heuristic that closely tracks the optimal
    	performance, and present a numerical performance evaluation for the
    	bounds and heuristic.
\end{enumerate}

The rest of this paper is organized as follows. In Section
\ref{sec:related_work} we survey the related work on network lifetime. In
Section \ref{sec:system_model}, we describe our system model and give
optimization formulations to solve for the maximum network lifetime.  Section
\ref{sec:numerical_study} details our numerical study and results for varying
network sizes. Finally, Section \ref{sec:conclusion} concludes this paper.

\section{Related Work}\label{sec:related_work}

In \cite{fitzgerald2018energy}, we considered the problem of
data aggregation and dissemination in IoT networks serving, for example,
monitoring, sensing, or machine control applications. A key aspect of the IoT
that differentiates it from classical wireless sensor networks (WSNs) is its
heterogeneity. We therefore considered cases where nodes may take on different
roles (for example, sensors, destinations, or transit nodes) for different
applications, and where multiple applications with different demands may be
present in the network simultaneously. Moreover, these demands can be more
general than only collecting data and forwarding it to a single sink, as is
usually the case for WSNs. Rather, data may be processed within the network (we
take the specific case of aggregation), and may be disseminated to multiple
sinks via multicast transmissions.

However, in that work, we focused on minimizing the total energy usage. We now
seek to extend this to consider the network lifetime. While total energy usage
may be important in, for example, green networking, in which we wish to reduce
the environmental impact and thus the overall energy usage, network lifetime is
a critical performance measure both for traditional WSNs and for emerging IoT
networks. Network lifetime gives a measure of how long the network can operate
without intervention and, in cases where it is impractical to charge nodes or
change their batteries, it gives the total operating time for the network.

Network lifetime has been studied extensively in the context of WSNs since the
early 2000's. A full review of the literature in this area is therefore beyond
the scope of this paper; a recent survey can be found in
\cite{yetgin2017survey}. We will instead focus on the recent work that is most
relevant to the current paper.

There are numerous different definitions of network lifetime adopted in the
literature \cite{yetgin2017survey}. Some of these include that the network
lifetime expires at the time instant a certain number (possibly as low as one)
or proportion of nodes deplete their batteries, when the first data collection
failure occurs, or when the specific node with the highest consumption rate runs
out of energy. In \cite{yetgin2017survey}, these definitions are classified into
four categories depending on whether they are based on node lifetime, coverage
and connectivity, transmission, or a combination of parameters.

However, a problem with many of these definitions is that they are not
application-centric. In practice, whether or not a network is functional
depends on the specific application or applications which it serves. Some
applications may require all nodes in the network to have remaining energy,
while others may continue to operate correctly with only a few nodes working.
The lifetime also depends on the capabilities of the network. For example, if
the network can be reconfigured, the lifetime may be extended by switching
configurations. This can be facilitated by the use of software defined
networking \cite{li2017survey}, as well as support from cloud services that
are capable of performing even demanding calculations to determine the best
network configuration at any given time, without incurring an energy cost in the
end devices.

This is the approach we adopt in this paper, and we define valid
configurations based on the demands of the applications present in the network
along with the roles the various nodes play in these demands. As such, we will
adopt a general definition of the network lifetime as the total time in which
the network is operational. Since we consider a class of applications with data
streams as their demands, this is most similar to the definition used in
\cite{cardei2005energy}, where the network lifetime was defined as the number of
sensory information task cycles achieved until the network ceases to be fully
operational.

There have been numerous techniques developed to improve the network lifetime in
specific use cases, mostly for WSNs consisting of homogeneous sensor nodes and a
single sink. In \cite{lin2018mpf}, the schedule and charge amounts of a mobile
vehicle that charges nodes were optimized, while in
\cite{mansourkiaie2017maximizing} nodes may regain energy through energy
harvesting, and routing is then optimized to maximize the lifetime. Routing is
also the focus of \cite{iova2015using}, however here an energy-balancing routing
protocol is developed, rather than determining optimal routes. Multilayer
optimization approaches are adopted in \cite{yetgin2015cross} and
\cite{keskin2014wireless}, covering multiple different aspects of WSN design. In
\cite{yetgin2015cross}, the design of the physical, medium access control, and
network layers was jointly optimized, including flow routing, link scheduling,
transmission rate selection, and node power allocation. This resulted in a
non-convex optimization problem that was difficult to solve, even for the simple
string topology considered. Meanwhile, in \cite{keskin2014wireless}, sensor
location, activity scheduling, sink mobility, and data routing were jointly
optimized. All of the above criteria were included directly in the mixed-integer
programming formulation, meaning that it is quite specific to the particular use
case considered. Sink placement for network lifetime maximization is
investigated in \cite{ahmed2018optimized} using k-nearest neighbour optimization
with the whale meta-heuristic.

In all of the above work, the nodes in the network are homogeneous, all
performing both sensing and data forwarding. No data processing is performed in
the network, and only a single configuration is used, rather than reconfiguring
the network in order to assist with energy balancing and thus extend the network
lifetime. Moreover, only a single application demand, consisting of collecting
data from all nodes to a single sink, can be accommodated. In this paper, we instead
optimize the network lifetime for a network that may host a general class of
heterogeneous application demands, and in which nodes may play different roles
and perform different operations for different applications.

Some work has been performed regarding network lifetime for networks with
heterogeneous nodes, but only in a quite limited sense. For example, there is
work based on the LEACH clustering protocol \cite{heinzelman2000energy,
heinzelman2002application}, where each node may be either an ordinary sensor
node or a cluster head at different times. Examples of variations on LEACH that
improve the network lifetime include \cite{anil2018novel},
\cite{nayak2016fuzzy} and \cite{leu2015energy}, while
\cite{osamy2018algorithm} presents a clustering routing protocol that considers
both network lifetime and coverage. In \cite{halder2014enhancement}, the nodes
are also heterogeneous, however they may only be of two types: sensor nodes and
relay nodes. This is also the case in \cite{yildiz2019maximization}, where
network lifetime is defined as the time until the first node depletes its
battery, and (unicast) routing is then optimised for each traffic flow to reach
the sink.

Some work in the literature also considers in-network processing. In
\cite{kale2019scheduling}, data aggregation trees are constructed and
scheduled, and the network can be reconfigured, in that different trees can
be used in different time periods. This work again uses the
traditional WSN model of many homogeneous sensor nodes all sending
measurements to a single sink. The scenario considered in
\cite{raptis2018maximizing} focuses on a machine-to-machine communication
application similar to the one we consider, including the presence of edge
nodes in the network. However, there, the problem addressed is that of data
placement on these edge nodes in order to maximize the network lifetime
under latency constraints. Routing is performed by selecting the paths that
yield the maximum lifetime, defined as the time until any node runs out of
energy; reconfiguration of the network as we propose in this paper is not
considered.

A few general frameworks for maximizing network lifetime have also been
developed. In \cite{cheng2008general}, the focus is on network deployment,
specifically the initial energy allocated to each node. Once again nodes are
homogeneous, with all nodes collecting data and transmitting it to their
neighbors, and the definition of network lifetime is the time until the first
sensor depletes its battery. A more general definition of network lifetime is
used in \cite{chen2005lifetime}, which applies a framework based on channel
states aimed at developing medium access protocols for improved lifetime.
However, nodes have fixed roles and only a single application is considered.

The most similar approach to our work can be found in \cite{castano2018exact},
where nodes may take on multiple different roles at different times. Indeed,
there, a similar solution method to the one we employ, based on column
generation, is used. However, in \cite{castano2018exact}, the pricing problem
for column generation requires enumeration of connected components in the
network graph and so is solved with the help of cut generation, whereas we
explicitly list constraints for valid routing trees in our pricing problem and
solve it directly. Further, there are a number of key differences in the problem
considered that differentiates our work here from that in
\cite{castano2018exact}. Firstly, only a single, specific monitoring application
is considered, and as such the network lifetime definition adopted is based on
coverage of the target area, rather than the more general definition we take.
The aim is then only to cover the targets and the interdependencies between
nodes required to establish valid routing trees are not considered. In fact,
cases where the traffic through the nodes has a significant impact on nodes'
power consumption is identified in \cite{castano2018exact} as a direction for
future work. This is exactly the case we address here, where applications
consist of data streams and as such transmission represents a major
energy-consuming operation for the nodes in the network.

\section{System Model}\label{sec:system_model}

We take as our starting point the scenario described in
\cite{fitzgerald2018energy}, that is a wireless multihop network carrying out
machine-to-machine communication. Within the network, some nodes are able to
act as sensors, collecting information about their environment, and some nodes
are actuators, able to use the collected sensor information and carry out
tasks. Nodes that are neither sensors nor actuators may transit data through
the network, possibly aggregating it along the way, and we refer to these nodes
as aggregators. As in \cite{fitzgerald2018energy}, a stream is defined as data
that is able to be aggregated. In this paper, we use the term (wireless) mesh
network to refer to the network topology of multiple wireless hops in a
non-hierarchical mesh, and machine-to-machine communication to refer to the
application performed by the network: communication between machines, which may
have sensors, actuators, or both.

In \cite{fitzgerald2018energy}, we considered two different data collection
models, however in this work we will focus on the second and more difficult of
these --- referred to as the $nK$ case --- in which $n$ different actuator nodes
must each collect sensor measurements from $K$ different sensor nodes. This use
case requires that data is both aggregated as it is collected from the sensor
nodes, and disseminated via multicast transmissions to multiple actuator nodes.

We then seek to maximize the network lifetime. We define network lifetime as the
time until the network is no longer able to carry out the above task, that is,
the time until $n$ different actuators are no longer able to each collect $K$
different sensor measurements. Here, $n$ may in general be smaller than the
total number of actuators, and $K$ may in general be smaller than the total
number of sensors. Moreover, the above definition does not specify how the
measurements should be aggregated, routed, and disseminated throughout the
network. In fact, we will allow the choice of sensor, actuator and aggregator
nodes, as well as the routing, to be varied during the network's operation in
order to extend its lifetime as some nodes deplete their batteries.

To this end, we define a network configuration as a set of chosen sensor,
aggregator, and actuator nodes, as well as appropriate routing to take
measurements from the sensor nodes, aggregate and transit them through the
network via the aggregator nodes, and then disseminate them to the actuator
nodes. In order to be valid, each network configuration must fulfill the
$nK$-condition of $n$ different actuator nodes each collecting $K$ different
measurements. Note that while each actuator node requires $K$ different
measurements, these may be common to multiple actuator nodes.

\begin{figure}
    \begin{subfigure}{0.48\columnwidth}
	\centering
	\includegraphics[width=\columnwidth]{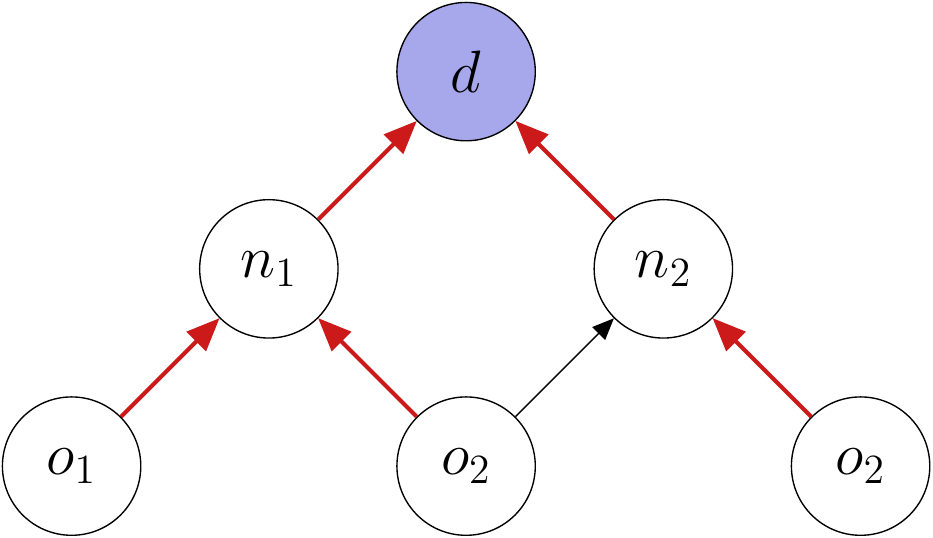}
	\caption{}
	\label{fig:example1}
    \end{subfigure}\hfill%
    \begin{subfigure}{0.48\columnwidth}
	\centering
	\includegraphics[width=\columnwidth]{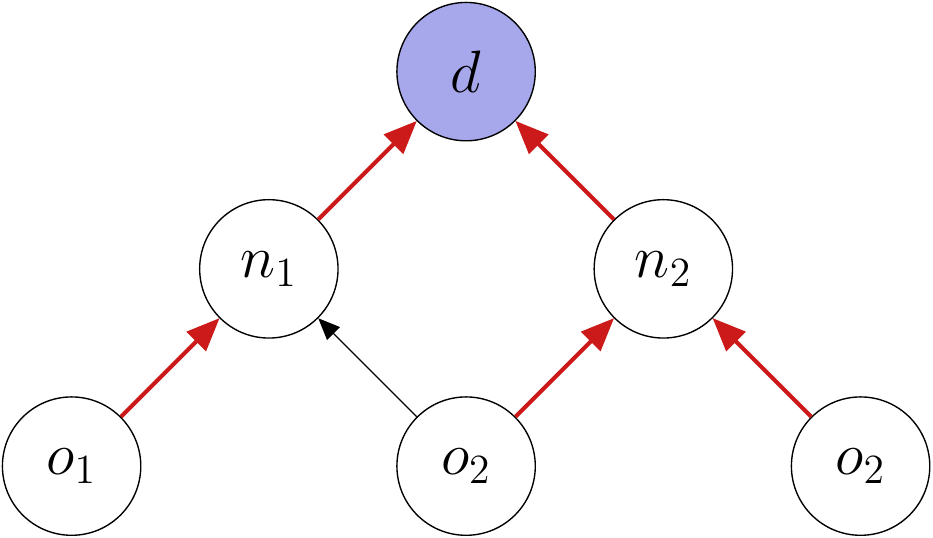}
	\caption{}
	\label{fig:example2}
    \end{subfigure}
    \caption{Two different network configurations to collect three different
    measurements at a single destination. Links used by each configuration are
    shown in red.}
    \label{fig:example}
\end{figure}

A simple example network is shown in Figure~\ref{fig:example}, with two different
configurations. The destination node, shown in the figure in blue and labelled
as $d$, must collect three different sensor measurements, that is one each from
origin nodes $o_1$, $o_2$, and $o_3$. The measurements from $o_1$ and $o_3$ must
be routed through aggregator nodes $n_1$ and $n_2$, respectively, since these
are the only available nodes in range. However, the measurement from $o_2$ may
be aggregated and transited through either $n_1$ or $n_2$, since both are in
range of $o_2$. One network configuration is then defined for each of these
options.

Since aggregating two measurements takes additional energy compared with only
transiting a single measurement, in the configuration in
Figure~\ref{fig:example1}, node $n_1$ will have a higher energy cost, while in
the configuration in Figure~\ref{fig:example2}, node $n_2$ will use more energy.
As there are an odd number of measurements to be collected, in this case it is
not possible to evenly share the energy cost between the two aggregator nodes
within a single configuration. Using two configurations, however, allows us to
do so, so that if we use each configuration for an equal amount of time, nodes
$n_1$ and $n_2$ will deplete their batteries at the same rate. Assuming their
initial battery capacities are equal, using two configurations thus allows us to
increase the network lifetime when compared with using only one of the two
configurations. In the latter case, one of the aggregators would deplete its
battery faster, leaving some unused energy in the battery of the other
aggregator. When using both configurations, all of the energy is used.

In our system model, data collection occurs in measurement periods of equal
duration. During each measurement period, the nodes in the network perform their
required tasks (sensing, aggregating, receiving, and transmitting) to fulfill
the $nK$-condition. After a node has performed all its allocated tasks, it may
sleep until the next measurement period. We thus consider that nodes only
consume energy to perform their tasks, while the energy consumed during sleep is
negligible. The duration of each measurement period must of course be long
enough to carry out the entire data collection and dissemination task. In many
applications, it is in fact considerably longer, with measurements being
collected perhaps every hour or day. While in some cases, such as factory
automation, the network may operate on a tight control loop requiring low delay
and thus short measurement periods, such applications are not typically energy
constrained and so are not the focus of this work.

The network lifetime is then expressed in measurement periods, that is, the
lifetime is equivalent to the number of times the network can collect and
disseminate the required sensor values before it is no longer able to meet the
$nK$-condition. To achieve a given lifetime, each network configuration operates
for a designated \emph{timeshare}: a number of measurement periods. We may use
the lifetime and timeshares in one of two ways. The first of these is
that, in the case of an existing network, and given a set of tasks the network
must perform, we can determine the longest possible operating time,
together with the configurations and their timeshares needed to achieve
it.

We may also consider the case of network deployment, where there is
again a given set of tasks, along with a target lifetime that the network must
achieve. The lifetime will in general be limited by the battery capacities of
the nodes: if the nodes start with more energy, they can operate for longer. The
problem for network deployment is them to correctly dimension the nodes' battery
capacities such that the target network lifetime will be achieved. In this case,
it is beneficial to allow the configuration timeshares to be fractional.  Of
course, in reality, the network would not be operated for a fraction of a
measurement period, since doing so does not provide a fraction of the utility of
that measurement period. Indeed, in such a case it may be that no measurements
reach their destinations in time at all. Nonetheless, allowing fractional
timeshares allows us to determine the maximum possible network lifetime for
given battery capacities of the nodes, and then scale this solution to the
desired lifetime by adjusting the battery capacities accordingly. For
example, we may double the battery capacity of all nodes, and consequently
double all configuration timeshares, giving double the network lifetime.

Scaling fractional timeshares in this way will provide the optimal lifetime for
the larger battery capacities, whereas, as we will see in our numerical results
in Section \ref{sec:numerical_study}, this is not the case if the original
solution was only allowed to include integer timeshares. Of course, it may not
be possible to achieve fractional timeshares exactly, but as the battery
capacities increase, we can achieve greater precision when rounding the
timeshares. We thus have a trade-off between the generality of the solution and
its exactness. A fractional solution can be applied at any scale to achieve a
desired lifetime, but an integer solution can be applied with no quantization
error.

\subsection{Notation and definitions}

We represent the wireless network as a directed graph $\mathcal G = (\mathcal V,
\mathcal A)$, where links (i.e., directed arcs) are established between nodes if
they are able to communicate with a satisfactory SNR in the absence of any
interference. We denote the set of arcs incoming to a node $v \in \mathcal V$ by
$\delta^-(v)$, and the set of arcs outgoing from $v$ by $\delta^+(v)$. The set
of nodes $\mathcal V$ is composed of three mutually disjoint subsets: the set of
sensor (origin) nodes $\mathcal O$, the set of aggregator nodes $\mathcal N$,
and the set of actuator (destination) nodes $\mathcal D$. Thus $\mathcal V =
\mathcal O \cup \mathcal N \cup \mathcal D$. Origin nodes generate, transit, and
aggregate packets; aggregator nodes transit and aggregate packets, but do not
generate them; and destination nodes can aggregate packets but do not transit
them, and therefore $\delta^+(d) = \varnothing,\,d \in \mathcal D$.

We define a network configuration as a set of tasks performed by the nodes in
the network. A configuration is valid if it is able to deliver $K$ unique
measurements to each of $n$ unique destination nodes. Each configuration
therefore includes a subgraph $\mathcal G' \subseteq \mathcal G,\,\mathcal G' =
(\mathcal V', \mathcal A')$, with $\mathcal V' \subseteq \mathcal V$ and
$\mathcal A' \subseteq \mathcal A$, describing which nodes participate in the
configuration, and the links along which measurements can traverse. Each
node in the configuration is assigned to perform operations that may consist of
transmission, reception, aggregation, and/or sensing (for origin
nodes). The set of all possible valid network configurations is denoted
$\mathcal C$.

All packets arriving at a sensor or aggregator node are aggregated and then
broadcast. Since we assume that a measurement period is significantly longer
than the time required to collect and disseminate all sensor measurements, each
transmitting node only needs to make a single transmission --- if need be, all
transmissions can be conducted in series to avoid interference and the
measurements will still be delivered within the measurement period. This
simplifies the energy calculations, since we only need to count one transmission
per node, and we do not need to consider transmission power adjustment to
compensate for interference.

Moreover, we do not need to explicitly account for the energy required to receive
packets. If a node receives a single packet, it must always then re-transmit it
(unless it is a destination node), and so the energy required for reception can
simply be included in the transmission energy cost. If a node receives multiple
packets, it always aggregates them, and so the extra energy required to receive
them can be included in the aggregation energy cost, which in our formulations
is proportional to the number of received packets minus one.

It is assumed that only the origin and aggregator nodes have limited battery
capacity and this limitation does not apply to destination nodes $v \in \mathcal
D$. This is because these nodes are either gateways collecting data, or actuator
nodes that perform other, most likely highly energy-demanding, tasks, and so the
energy required for data reception and aggregation is not significant for these
nodes.\footnote{It would however not be difficult to modify our formulations to
    consider limited energy at destination nodes, if desired, by changing the
indexing sets for the energy constraints in the pricing problem.} Hence, each
node $v \in \mathcal O \cup \mathcal N$ has a specified (limited) battery
capacity $B(v)$, expressed in Joules (J), and in each network configuration $c
\in \mathcal C$, node $v$ uses $P(v,c)$ J of energy per measurement period.
$P(v,c)$ is thus the energy cost for node $v$ to deliver one entire set of
measurements fulfilling the $nK$-condition when configuration $c$ is active. For
each configuration $c \in \mathcal C$, we define the timeshare $t_c$ of $c$ to
be the number of measurement periods in which $c$ is scheduled to be active.

A summary of notation used is shown in Table~\ref{tab:notation}. Observe that in
our notation indices of a given parameter (if any) are put in brackets (like in
$P(v,c)$), while indices of variables are placed as subscripts and/or
superscripts (like in $z^{od}_a$). This convention, used for example in
\cite{korte12}, helps to make problem formulations readable.

\begin{table}
    \begin{centering}
	{\small
	\begin{tabularx}{\columnwidth}{|c|X|}
	    \hline
	    $\mathcal V$ & set of nodes (vertices) in the network \\
	    \hline
	    $\mathcal A$ & set of arcs $(v, w)$, $v, w \in \mathcal V$
	    indicating node $w$ is within transmission range of node $v$
	    (barring any interference)\\
	    \hline
	    $\mathcal O$ & set of origin (sensor) nodes\\
	    \hline
	    $\mathcal N$ & set of aggregator nodes\\
	    \hline
	    $\mathcal D$ & set of destination (actuator) nodes \\
	    \hline
	    $\mathcal C$ & set of network configurations \\
	    \hline
	    $\delta^-(v)$ & set of incoming arcs to node $v$ \\
	    \hline
	    $\delta^+(v)$ & set of outgoing arcs from node $v$ \\
	    \hline
	    $t_c$ & timeshare of network configuration $c \in \mathcal C$ \\
	    \hline
	    $x^{od}$ & whether or not the measurement from origin $o \in \mathcal O$ is
	    received by destination $d \in \mathcal D$ \\
	    \hline
	    $z^{od}_a$ & flow of the measurement from origin $o \in \mathcal O$
	    to destination $d \in \mathcal D$ on arc $a \in \mathcal A$ \\
	    \hline
	    $y^o_a$ & whether or not arc $a \in \mathcal A$ carries the
	    measurement from origin $o \in \mathcal O$ \\
	    \hline
	    $Y_a$ & whether or not arc $a \in \mathcal A$ carries an
	    (aggregated) measurement \\
	    \hline
	    $X^{o o'}_v$ & whether or not the measurements from origins
	    $o,\,o'\in \mathcal O$ are aggregated at node $v \in
	    \mathcal N \cup \mathcal D$ \\
	    \hline
	    $G_v$ & energy required to broadcast from node $v \in \mathcal O
	    \cup \mathcal N$ \\
	    \hline
	    $g_{v}$ & number of (aggregated) measurements aggregated at node
	    $v \in \mathcal O \cup \mathcal N$ minus $1$ (and $0$ if there is no
		aggregation at $v$) \\
	    \hline
	    $u^{o}$ & whether or not the measurement from origin
	    $o \in \mathcal O$ is received by any destination \\
	    \hline
	    $\mathcal O^{|2|}$ & set of all 2-element subsets of $\mathcal O$ \\
	    \hline
	    $T(a)$ & transmission energy required on arc $a$ \\
	    \hline
	    $S(v)$ & processing energy required for aggregation by node $v$ \\
	    \hline
	    $B(v)$ & battery capacity of node $v$ (in J)\\
	    \hline
	    $P(v,c)$ & energy used per measurement period by node $v$ in
	    configuration $c$ (in J)\\
	    \hline
	    $E(v, c)$ & battery depletion fraction per measurement period of node $v$ in configuration $c$\\
	    \hline
	    $\mathbb B$ & set of binary numbers $\{0, 1\}$ \\
	    \hline
	    $\mathbb R$ & set of real numbers \\
	    \hline
	    $\mathbb R_+$ & set of non-negative real numbers \\
	    \hline
	    $\mathbb Z_+$ & set of non-negative integers\\
	    \hline
	    \end{tabularx}
	}
	\caption{Summary of notation.}
	\label{tab:notation}
    \end{centering}
\end{table}

\subsection{Master problem}\label{sec:master_problem}

The network lifetime problem can be formulated as the following integer
programming problem, called the\emph{ master problem}:
\begin{subequations} \label{form:master_absolute}
    \begin{align}
	\mathrm{max} & \sum_{c \in \mathcal C} t_c & \label{ma_objective} \\
		& \sum_{c \in \mathcal C} t_c P(v, c) \leq B(v), & v \in \mathcal O \cup \mathcal N \label{ma_constraint}\\
	& t_c \in \mathbb Z_+, & c \in \mathcal C. \label{ma_variables}&
    \end{align}
\end{subequations}

The objective \eqref{ma_objective} maximizes the sum of the times, expressed in
measurement periods, in which each network configuration is active, giving the
total operating time for the network. Constraint \eqref{ma_constraint} requires
that the energy used by node $v \in \mathcal O \cup \mathcal N$ across all configurations does
not exceed $v$'s battery capacity. Constraint \eqref{ma_variables} specifies
that the timeshare allocated to each network configuration must consist of an
integer number of measurement periods. We thus find an exact optimal solution
for the given battery capacities.

For a solution that is
scalable with the battery capacities, but that may introduce quantization error
in realizing the timeshares, we can instead take the linear relaxation of
formulation \eqref{form:master_absolute}, that is, changing constraint
\eqref{ma_variables} to $t_c \in \mathbb R_+$, so that integrality of variables
$t_c, \, c \in \mathcal C,$ is relaxed. This linear relaxation is an important
element in our optimization approach to network lifetime maximization, and is
formulated as follows.
\begin{subequations} \label{form:LR}
    \begin{align}
	\mathrm{max} & \sum_{c \in \mathcal C} t_c \label{form:LR-1} \\
	    & \sum_{c \in \mathcal C} t_c E(v, c) \leq 1, & v \in \mathcal O \cup \mathcal N \label{form:LR-2}\\
	    & t_c \in \mathbb R_+, & c \in \mathcal C,
    \end{align}
\end{subequations}
where $E(v, c) = \frac{P(v, c)}{B(v)}, \, v \in \mathcal O \cup \mathcal N, c
\in \mathcal C,$ defines the (dimensionless) \emph{depletion fraction} of the
battery of node $v$, that is, the proportion of node $v$'s total battery
capacity that is used up during one measurement period when configuration $c$ is
applied.  Formulation \eqref{form:LR} is obtained from
\eqref{form:master_absolute} by dividing both sides of inequalities
\eqref{ma_constraint} by $B(v)$ (which is assumed to be greater than $0$).
Observe that if the values $t^*_c, \, c \in \mathcal C,$ form an optimal
solution of the linear relaxation \eqref{form:LR} then $\lfloor t^*_c \rfloor,
\, c \in \mathcal C,$ constitute a feasible solution of the master problem
\eqref{form:master_absolute}. Moreover, such a solution tends to be close to
optimal when the lifetime of the network consists of a large number of
measurement periods, that is, when nodes' depletion fractions are low.

It is important to observe that the master problem formulated here is
non-compact, which means that it has an exponential number of variables $t_c, \,
c \in \mathcal C,$ since in general the number of valid network configurations
(i.e., $|\mathcal C|$) grows exponentially with the size of the network. We will
come back to this issue in Section~\ref{subsec:pricing}.

\subsection{Dual problem} \label{subsec:dual}

In order to solve the master problem above, we first solve its linear relaxation
\eqref{form:LR} by column generation \cite{Lasdon}. We start with an initial set
of network configurations $\mathcal C'$ (where $\mathcal C' \subset \mathcal C$)
and then iteratively generate new configurations that can improve the objective
\eqref{form:LR-1}. To do this, we first need to take the dual \cite{Lasdon} of
the linear programming problem represented by \eqref{form:LR} with $\mathcal C$
substituted with $\mathcal C'$ (called the primal problem in this context). The
dual problem is as follows:
\begin{subequations} \label{form:dual_relative}
    \begin{align}
	\mathrm{min} & \sum_{v \in \mathcal O \cup \mathcal N} \pi_v & \\
	    & \sum_{v \in \mathcal O \cup \mathcal N} \pi_v E(v, c) \geq 1, & c \in \mathcal C' \label{form:dual_relative-2}\\
	    & \pi_v \in \mathbb R_+, & v \in \mathcal O \cup \mathcal N,
    \end{align}
\end{subequations}
where $\pi_v, \, v \in \mathcal O \cup \mathcal N,$ are dual variables corresponding to the primal constraints \eqref{form:LR-2}.
Note the nice symmetry exhibited by the primal and dual problems, with the role of network configurations and timeshares interchanged.

\subsection{Pricing problem} \label{subsec:pricing}

To generate new improving configuration (if any) we need a \emph{ pricing
problem}, and this is where the main complexity lies. The master problem itself
is very general and could apply to any type of energy-draining task in which the
nodes deplete their batteries at different rates in different configurations.
However, it is the pricing problem that finds a proper improving configuration
$c'$ (among all valid configurations in $\mathcal C \setminus \mathcal C'$) and
delivers the resulting depletion fractions $E(v,c')$ implied by $c'$ to be used
in formulation \eqref{form:dual_relative} with $\mathcal C'$ augmented with
$c'$.  Such a pricing problem for the $nK$ use case is shown in formulation
\eqref{form:pp}, and is based on the $nK$ formulation in
\cite{fitzgerald2018energy}\footnote{We show here the formulation for a single
    application data stream. However, this can be adapted to multiple data
    streams in a straightforward way by adding indices $s$, for $s$ in the
    set of data streams, to both the variables and node sets. This allows the
set of available origin, aggregator, and destination nodes to be specific to
each stream. See \cite{fitzgerald2018energy} for more details.}. Note that the
pricing problem makes use of an optimal solution $\pi^*_v, \, v \in \mathcal O
\cup \mathcal N,$ of the dual.
\begin{subequations} \label{form:pp}
   \begin{align}
       \min \quad & \sum_{v \in \mathcal O \cup \mathcal N} \pi_v^*
   E_v & \label{pp:objective} \\
	& \sum_{o \in \mathcal O} x^{od} \ge K, & d \in \mathcal D \label{nk:data_collection} \\
	& \sum_{a \in \delta^+(v)} z^{od}_a = \sum_{a \in \delta^-(v)}
       z^{od}_a, &  o \in \mathcal O, \, d \in \mathcal D, \, v \in \mathcal V \setminus \{o,d\} \label{nk:flow in transit nodes} \\
       & \sum_{a \in \delta^-(d)} z^{od}_a = x^{od}, & o \in \mathcal O, \, d \in \mathcal D \label{nk:flow to destination} \\
    & z^{od}_a \le Y_a, & o \in \mathcal O, \, d \in \mathcal D, \, a \in \mathcal A \label{nk:flow feasibility} \\
    & Y_a \leq \sum_{o \in \mathcal O} \sum_{d \in \mathcal D}
		 z_a^{od}, & a \in \mathcal A \label{nk:no_flow_on_arc} \\
    & z^{od}_a \le y^o_a, & o \in \mathcal O, \, d \in \mathcal D, \, a \in \mathcal A \label{nk:link_usage-1} \\
    & y^o_a \le \sum_{d \in \mathcal D} z^{od}_a , & o \in \mathcal O, \, a \in \mathcal A \label{nk:link_usage-2} \\
    & \sum_{a \in \delta^-(v)} y^o_a \le 1, & o \in \mathcal O, \, v \in \mathcal V \label{nk:no_double-aggregation-1} \\
    & X^{o o'}_v \ge y^{o}_a + \big( \sum_{a' \in \delta^-(v)
    \setminus \{ a \}} y^{o'}_{a'} \big) - 1, & v \in \mathcal V, a \in \delta^-(v), \, \{o, o'\} \in \mathcal O^{|2|} \label{nk:no_double-aggregation-2-new1} \\
    & \sum_{v \in \mathcal V} X^{o o'}_v \le 1, & \{o, o'\} \in \mathcal O^{|2|} \label{nk:no_double-aggregation-3} \\
    & g_{v} \geq \sum_{a \in \delta^-(v)} Y_a - 1 , & v \in \mathcal N \label{nk:at_least_two} \\
      & g_{o} \geq \sum_{a \in \delta^-(v)} Y_a + u^o - 1 , & o \in \mathcal O \label{nk:at_least_two_origins} \\
      & u^o \geq x^{od}, & o \in \mathcal O, \, d \in \mathcal D \label{nk:origin_used} \\
      & G_{v} \geq T(a)Y_a , & v \in \mathcal O \cup \mathcal N, \, a \in \delta^+(v) \label{nk:node_activity} \\
      & E_v = \frac{G_{v} + S(v) g_v}{B(v)}, & v \in \mathcal O \cup
    \mathcal N \label{nk:energy_calc}\\
	& x^{od} \in \mathbb B, &  o \in \mathcal O, \, d \in \mathcal D \\
	& z^{od}_a \in \mathbb B, & o \in \mathcal O, \, d \in \mathcal D, \, a \in \mathcal A \\
    & y^o_a \in \mathbb R_+, & o \in \mathcal O, \, a \in \mathcal A\\
	& Y_a \in \mathbb R_+, & a \in \mathcal A\\
 & X^{o o'}_v \in \mathbb R_+, & v \in \mathcal N \cup \mathcal D,
    \, \{o, o'\} \in \mathcal O^{|2|}\\
	& u^o \in \mathbb B, & o \in \mathcal O \\
 & g_{v} \in \mathbb R_+, & v \in \mathcal V\\
 & G_v, E_v \in \mathbb R_+, & v \in \mathcal O \cup \mathcal N.
    \end{align}
\end{subequations}

The pricing problem generates a network configuration that performs the task of
delivering $K$ unique measurements to each of $n$ destinations. It must therefore
ensure that the routing used for the measurements is correct --- that is, each
measurement follows a non-cyclic path to each destination to which it is
delivered. Further, it makes sure that whenever a node receives more than one
measurement, it aggregates them, transmitting only a single, aggregated packet
further along the route. Since the goal of our pricing problem is not only to find
a valid configuration, but to find a configuration that improves the network
lifetime as much as possible, it must also consider the energy used by the nodes
in performing their tasks in the configuration. To this end, the pricing problem
calculates the energy needed by each node for both transmission and aggregation.

The first decision variable in the pricing problem, $x^{od}$, will be set to $1$
if the measurement collected by origin node $o \in \mathcal O$ is delivered to
destination node $d \in \mathcal D$. Constraint \eqref{nk:data_collection} then
guarantees the $nK$-condition, that is, that each selected destination node
receives at least $K$ measurements.  The next set of variables and constraints
concern routing. Variables $z_a^{od}$ describe flows from origin node $o \in
\mathcal O$ towards destination node $d \in \mathcal D$ along arc $a \in
\mathcal A$. Constraints \eqref{nk:flow in transit nodes} and \eqref{nk:flow to
destination} then provide flow conservation, subject to destination $d$ being
selected to collect a measurement from origin $o$ ($x^{od} = 1$). The variable
$Y_a$, $a \in \mathcal A$, will be $1$ if arc $a$ is used to carry any flow, and
$0$ otherwise. This is ensured by constraints \eqref{nk:flow
feasibility}--\eqref{nk:link_usage-2}. Although variables $Y_a$ are formally
continuous, in the optimal solution they will only take the values $1$ or $0$,
since they are forced to $1$ by binary variables $z_e^{od}$ on arcs used for
transmission (constraint \eqref{nk:flow feasibility}), while on arcs with no
transmissions they are forced to $1$ (constraint \eqref{nk:no_flow_on_arc}).
Variables $y^o_a$ describe whether or not arc $a \in \mathcal A$ is used to
carry the measurement from origin node $o \in \mathcal O$, and have the same
property of being set to only $1$ or $0$ in the optimal solution, since they are
either forced to $1$ by constraint \eqref{nk:link_usage-1}, or to zero by
constraint \eqref{nk:link_usage-2}.

The next part of the pricing problem concerns aggregation. Firstly, constraint
\eqref{nk:no_double-aggregation-1} prevents a node from receiving a given
measurement on more than one arc. Variable $X_v^{oo'}$ records where aggregation
occurs for each pair of measurements; it will be $1$ if and only if the
measurements from origin nodes $o$ and $o'$, $o, \, o' \in \mathcal O$, are
aggregated at node $v \in \mathcal V$. If a node receives measurements from two
different origins on different arcs, constraint
\eqref{nk:no_double-aggregation-2-new1} then forces the node to aggregate the
measurements. Lastly, constraints
\eqref{nk:no_double-aggregation-3} make sure that two packets from different
origin nodes can be aggregated at most once.  

The remaining constraints and variables are used to calculate the energy costs.
Any node that receives at least two packets aggregates them, and will incur a
processing cost proportional to the number of packets aggregated less one.
Constraints \eqref{nk:at_least_two}--\eqref{nk:origin_used} calculate the number
of aggregation operations for each node $v \in \mathcal V$, and record this in
variable $g_v$. Here, a special case occurs for origin nodes selected to provide
measurements, as each such node aggregates an extra packet (its own
measurement). This is indicated by variable $u^o$. If a node transmits a packet,
this also carries an energy cost, placed in variable $G_v$ by constraint
\eqref{nk:node_activity}. Each node's transmission energy cost is given by the
highest transmission cost $T(a)$, $a \in \mathcal A$, for any arc on which it
transmits. 

Finally, the depletion fraction for each node $v \in \mathcal V$ is computed in
constraint \eqref{nk:energy_calc} and placed in variable $E_v$. Here, the total
aggregation cost is given by the term $S(v)g_v$, where $S(v)$ is the energy cost
for each aggregation operation and, as previously mentioned, $g_v$ gives the
number of aggregation operations performed by node $v$. This is added to $v$'s
transmission energy cost $G_v$ and divided by its battery capacity $B(v)$ to
give the depletion fraction.  Aside from the energy calculation in constraint
\eqref{nk:energy_calc}, which is modified to represent each node's depletion
fraction instead of its absolute energy usage, the constraints for the pricing
problem are the same as for the total energy minimization problem for the $nK$
use case as defined in \cite{fitzgerald2018energy}. Otherwise, it is only the
objective that needs to be changed to reflect the dual constraint
\eqref{form:dual_relative-2}.

The pricing problem generates a network configuration $c'$ with $E(v,c') =
E^*_v$, where $E^*_v, \, v \in \mathcal O \cup \mathcal N,$ is an optimal
solution of \eqref{form:pp}. The generated configuration $c'$ is an improving
configuration (that is added to the current set of configurations $\mathcal C'$)
only when the resulting optimal objective \eqref{pp:objective} is strictly less
than $1$. This is because adding constraint \eqref{form:dual_relative-2}
corresponding to $c'$ to the dual formulation \eqref{form:dual_relative} will
make the current optimal dual solution $\pi^*_v, \, v \in \mathcal O \cup
\mathcal N,$ infeasible. Moreover, the constraint generated by the new
configuration is violated by the optimal dual solution in question to the
maximal extent; in fact, the value of this violation is equal to what is usually
called the reduced cost of the non-basic variable $c'$ in the simplex algorithm.
On the other hand, when the optimal objective is greater than or equal to $1$,
there is no improving configuration outside $\mathcal C'$ and therefore
$\mathcal C'$ is sufficient to solve the linear relaxation \eqref{form:LR} to
optimality even though not all configurations in $\mathcal C$ are directly
considered.

The strength of our solution approach can be seen here. We use a general
framework for solving for the network lifetime, with the specifics of the
task(s) the network is to perform relegated to the pricing problem. The
complexity and difficulty of properly formulating constraints for routing and
aggregation as in our pricing problem is typical of many other network problems.
By adopting our approach, the lifetime can be maximized for any task, and the
formulation of the specific task constraints is a relatively independent
undertaking, with only the objective determined by the dual problem
\eqref{form:dual_relative}.

\subsection{Solving the master problem} \label{sec:solve-MP}

For solving the (integer) master problem formulated in
\eqref{form:master_absolute} we use the so-called price-and-branch (P\&B)
two-stage algorithm \cite{pioro2012network}. In the first stage we solve the
linear relaxation \eqref{form:LR} of the master problem by column generation
that involves, as explained above, solving the pricing problem \eqref{form:pp}
(that is why the word ``price'' appears in P\&B). Then, in the second stage, we
solve formulation \eqref{form:master_absolute} through the standard
branch-and-bound (B\&B) algorithm (that is why the word ``branch'' is used in
P\&B) available in mixed-integer programming solvers, such as CPLEX, for the
fixed set $\mathcal C'$ of configurations resulting from the column generation
algorithm. Clearly, the so obtained solution of the master problem is in general
suboptimal as there is no guarantee that the set $\mathcal C'$ contains a subset
of the configurations necessary to achieve the optimum (which would be
guaranteed if the set $\mathcal C$ of all configuration were applied).

Actually, to assure true optimality, the master problem should be solved using
the branch-and-price (B\&P) algorithm \cite{pioro2012network} instead of P\&B.
The basic difference between B\&P and P\&B is that in the latter the column
generation algorithm is invoked only once, at the root node of the B\&B tree,
and then the linear subproblem solved at each of the subsequent B\&B nodes
assumes the subfamily $\mathcal C'$ computed at the root. B\&P in turn, would
apply the column generation algorithm at each B\&B node. Because of this, B\&P
would consume excessive overall computational time even for medium size networks.

In fact, it is also possible to solve the master problem to optimality using a
compact mixed-integer problem formulation (instead of using the non-compact
formulation \eqref{form:master_absolute} together with the pricing problem
\eqref{form:pp}) for which the number of variables and constraints is polynomial
in the size of the network and the battery capacity. In such a formulation, the
configurations for the consecutive measurement periods are specified explicitly
by means of additional binary variables and corresponding constraints (for each
measurement period) in the way used in the pricing problem. The so obtained
formulation could be solved directly, using a mixed-integer programming solver,
but this would involve a number of binary variables that is far beyond the reach
of current solvers. For this reason, we take the more practical approach of
P\&B to solve this computationally hard problem.

As already observed in Section~\ref{sec:master_problem}, a feasible solution of
the master problem can be easily obtained by rounding down the optimal values
$t^*_c, \, c \in \mathcal C',$ of the linear relaxation resulting from the
column generation algorithm. The quality of such an integer solution can in
general be improved by solving the master problem for the set of configurations
$\mathcal C''$, where $\mathcal C'' = \{ c \in \mathcal C': \, t^*_c > 0 \}$.
This may considerably decrease the number of variables in
\eqref{form:master_absolute} (and thus speed up the computations) since the
number of configurations in the set $\mathcal C''$ is not greater than the
number of nodes with entirely exhausted batteries in the optimal solution of the
linear relaxation --- this follows from the form of the basic optimal solution
of a linear programming problem \cite{Minoux-book}. Such an obtained solution
could also be used as an initial lower bound in the second stage of the P\&B
algorithm, improving its performance.

\subsection{Practical Implementation}

As our results in Section~\ref{sec:numerical_study} will show, the time needed
to solve our optimization formulations using the approach described above is
feasible for practical implementation of this approach provided that the network
lifetime obtained is sufficiently long. We will defer further discussion of the
solution times to Section~\ref{sec:numerical_study}, however there are a number
of other issues that need to be addressed to realise a practical deployment.
These are how and where the optimization is performed, how the nodes in the
network are informed of the configurations to use and their roles in them, and
what signalling is required to initiate each reconfiguration of the network at the
correct time.

In many networks performing machine-to-machine communication, a connection to
the wider Internet is present in at least some nodes. In fact, this can be
regarded as the typical case, and increasingly so in future, as new radio
technologies such as LPWAN and 5G bring Internet connectivity to more areas. In
this case, the optimization problems can easily be solved in the cloud, with its
abundant computing resources, and the results communicated to the nodes in the
mesh network via the Internet gateways. However, even in the case of a
standalone network, since the configurations can be computed in advance, it is
possible to solve the optimization problems before deployment of the network,
with the results then pre-programmed into the nodes. This adds to the storage
requirements of the nodes, since they must store the timing of each
configuration and their role in it, however this increase is modest since, as we
will see in Section~\ref{sec:numerical_study}, only few configurations are
needed to reach the maximal lifetime.

A more difficult issue is that of coordinating the nodes to perform the actual
reconfigurations. Correct updating of network flow routing is a non-trivial
issue that has been the subject of much research, both in traditional IP
networks and software-defined networks \cite{li2017survey}, with potential
pitfalls such as forwarding loops and forwarding black holes if nodes are
updated in the wrong order. As such, the definition of a protocol to ensure
correct operation of the network during reconfiguration is beyond the scope of
this work, but existing work on software-defined network updating could be used as
a basis for this.

Nonetheless, it is clear that synchronization is required so that all nodes will
update their configurations at the designated time. However, since each
configuration is expected to be used for at least hours, and more likely weeks
or longer, this synchronization does not need to be particularly precise. One
possible mechanism could be the designation of one or more controller nodes in
the network that can disseminate a reconfiguration message to the other nodes,
for example via simple flooding, when it is time to adopt the next
configuration. Relative to the time of operation of each configuration, this
will incur only a small overhead in network capacity and energy usage. In the
case of a network with an Internet connection, the controller may even be
external, placed in the cloud or a fog node.

\section{Numerical study}\label{sec:numerical_study}

We conducted a numerical study in which we generated networks using
\cite{network_generator}, with 10 to 30 nodes. The networks were generated using
the same methodology and parameters as in \cite{fitzgerald2018energy} in order
to have comparable results. Nodes were placed uniformly randomly in a square
area, with the area, number of measurements to collect, and number of sensor and
destination nodes all scaled with the network size. The exact parameters used
are given in Table \ref{tab:params}. For each network size, 20 different
networks were generated, and all results are presented with 95\% confidence
intervals over the different network instances.

\begin{table}
    \begin{centering}
	\begin{tabular}{|c|c|c|c|c|c|}
	\hline
	Nodes & Area width [m] & $K$ & $|\mathcal O|$ & $|\mathcal D|$ &
	$|\mathcal N|$ \\
	\hline
	10 & 122.47 & 3 & 4 & 2 & 4 \\
	\hline
	15 & 150.0 &  5 & 6 & 3 & 6 \\
	\hline
	20 & 173.21 & 6 & 8 & 3 & 9 \\
	\hline
	25 & 193.65 & 8 & 10 & 4 & 9 \\
	\hline
	30 & 212.13 & 9 & 12 & 5 & 13 \\
	\hline
	\end{tabular}
	\caption{Parameters used for the numerical study.}
	\label{tab:params}
    \end{centering}
\end{table}

In order to initialize the column generation, we used the total energy
minimization problem from \cite{fitzgerald2018energy}. This gives an initial
network configuration that minimizes the sum of the energy used by all nodes in
one measurement period. We then iterated through the column generation
process, solving first the linear relaxation of the master problem (formulation
\eqref{form:LR}) to get the optimal dual solution vector $\pi^*$, and then the
pricing problem to generate new network configurations. When column generation
was complete, the master problem was solved a final time to get
the optimal timeshares for each network configuration.

As discussed in Section~\ref{sec:master_problem}, at this point we could
either solve the master problem as an integer problem, yielding an exact
solution for the specific battery capacities given, or we could again
solve its linear relaxation, giving a solution that scales with the battery
capacities, albeit with a quantization error that depends on the absolute
capacities. In our numerical study, we solved both variants in order to compare
them. Since only the relative energy costs of each operation are needed, we set
the aggregation cost to 1 and the transmission cost to 5 as in
\cite{fitzgerald2018energy}. For the linear relaxation, we used a battery
capacity of 100, that is, a node may perform 100 aggregation operations or 20
transmission operations before depleting its battery. Here the actual capacity
is not important, as long as it is large enough to allow the network to complete
at least one measurement period. For the integer problem, however, we used a
battery capacity of 1000 so that we were able to accommodate a reasonable number
of different configurations in the solutions.

\subsection{Results}\label{sec:results}

\subsubsection{Linear Relaxation}

\begin{figure}
    \centering
    \includegraphics[width=0.5\columnwidth]{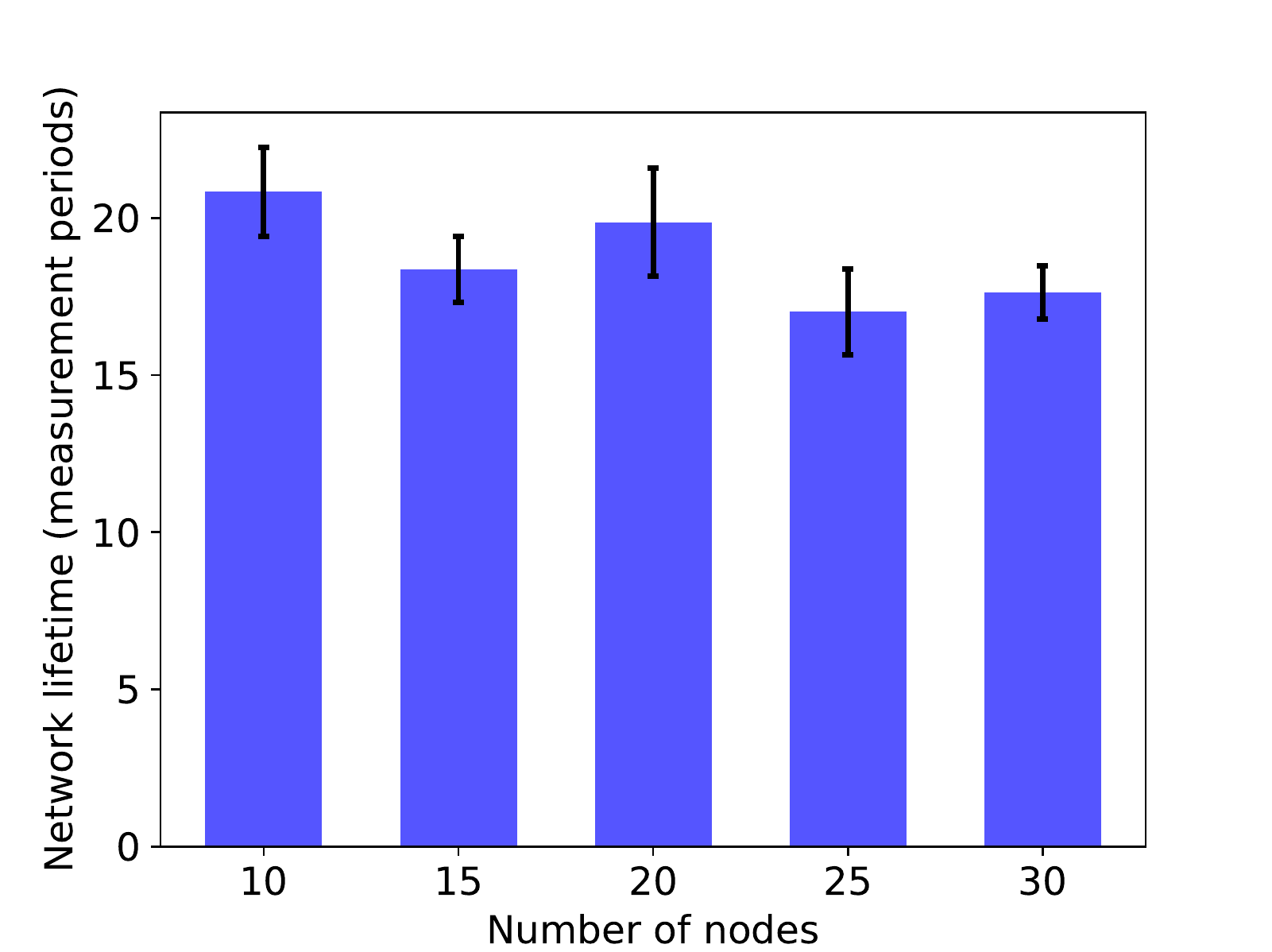}
    \caption{Network lifetime vs. number of nodes in the network (linear relaxation).}
    \label{fig:lifetime}
\end{figure}

Figure~\ref{fig:lifetime} shows the network lifetime vs. the network size, that
is, the number of nodes in the network, for the linear relaxation case. The
total network lifetime was quite consistent, with small variance for each
network size, and not much difference in lifetime across different network
sizes. However, there is a slight decrease in the lifetime as the number of
nodes increases. In \cite{fitzgerald2018energy}, both the total energy cost and
the max-min per node energy cost increased with the number of nodes, and the
lifetime reflects this same trend: with increasing energy costs, the lifetime
decreases.

\begin{figure}
    \centering
    \includegraphics[width=0.5\columnwidth]{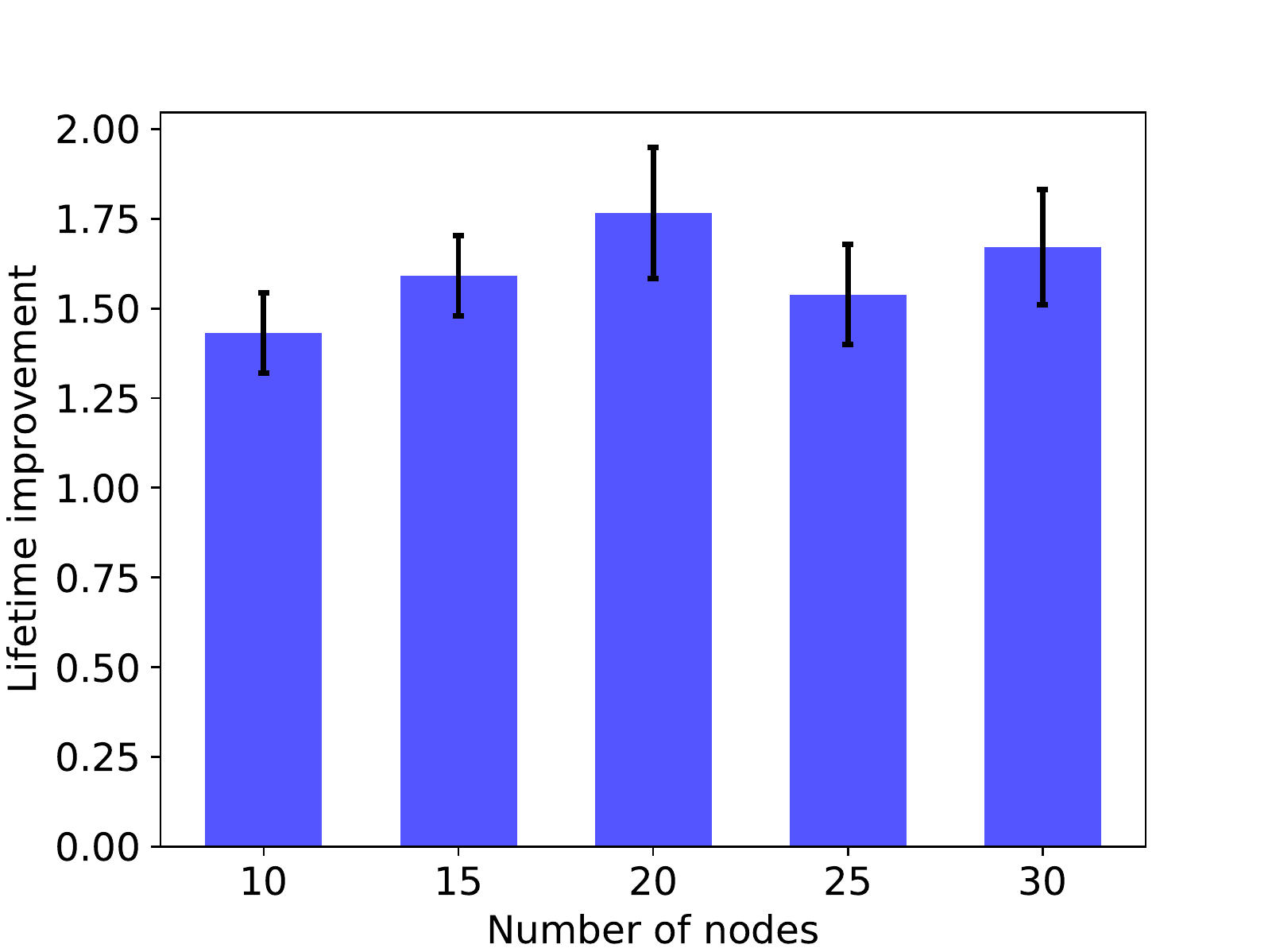}
    \caption{Network lifetime improvement compared with minimum total energy
    configuration vs. number of nodes in the network (linear relaxation).}
    \label{fig:lifetime_improvement}
\end{figure}

However, if we compare the improvement gained by optimizing for maximal network
lifetime, as opposed to for minimal total energy
(Figure~\ref{fig:lifetime_improvement}), we see that the improvement is
relatively flat across different network sizes. The largest improvement achieved
for the networks tested was approximately 1.75 times the lifetime when simply
using the configuration that gives the minimal total energy. Solution times
(averaged across all experiment runs)  for both the primal and dual problems
were very short: less than 0.01 s in all cases.

\begin{figure}
    \centering
    \includegraphics[width=0.5\columnwidth]{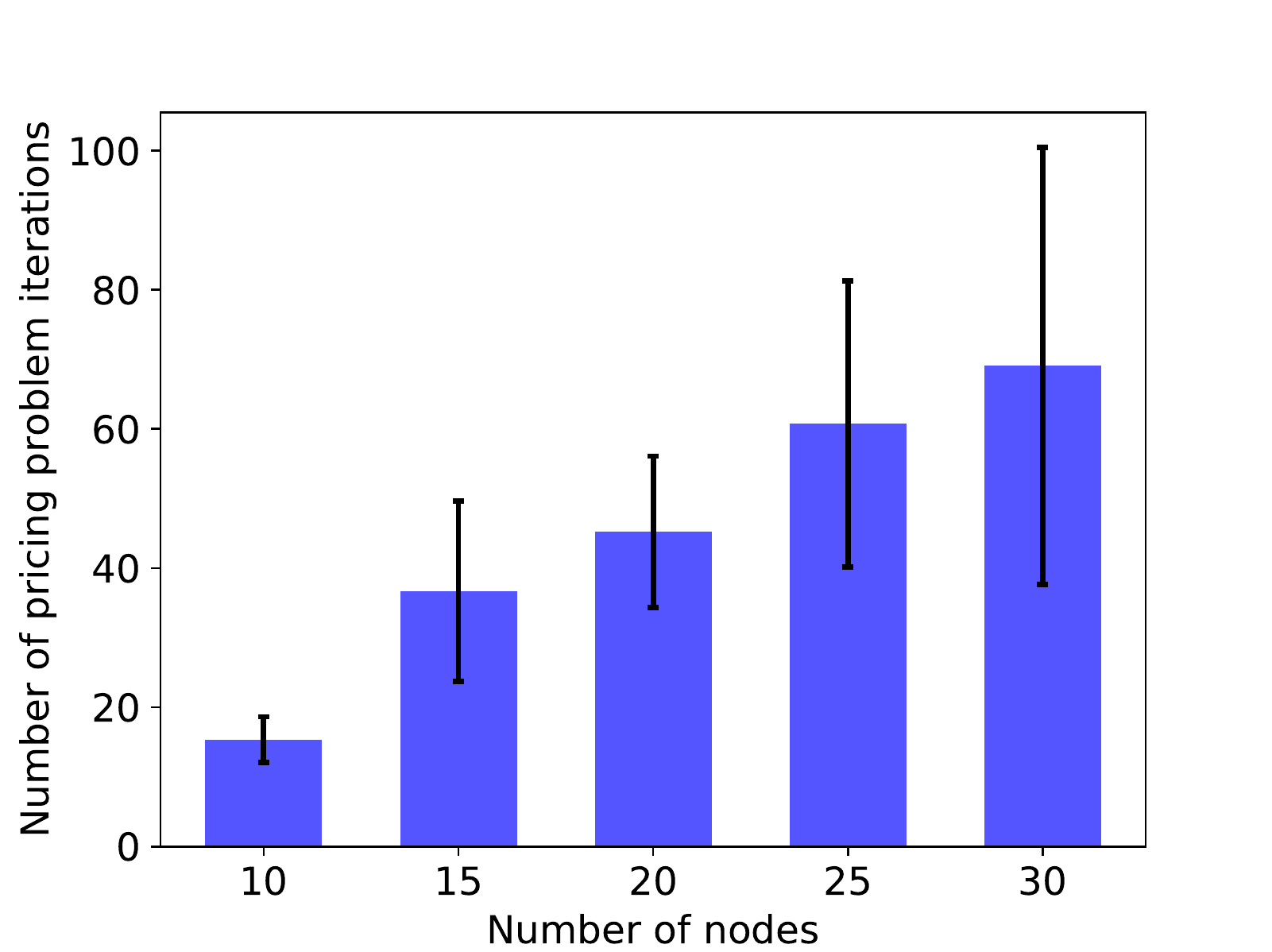}
    \caption{Pricing problem iterations needed to reach the final optimal solution (linear relaxation).}
    \label{fig:iterations}
\end{figure}

However, the solution times for (all iterations of) the pricing problem were
much longer: 1.6 s for 10 nodes and increasing exponentially to 817\,649 s (227
hours) for 30 nodes. As can be seen in Figure~\ref{fig:iterations}, this
increase is partly due to an increase in the number of iterations of the pricing
problem that are needed as the network size grows. The increasing solution times
indicate that as the network lifetime increases, the absolute battery capacity
of the nodes needs to be sufficiently large to make it worthwhile to obtain
optimal solutions for maximum lifetime. For example, if the network would
operate for multiple years --- not unreasonable in many IoT applications, for
example infrastructure monitoring --- then even long solution times for
optimization can easily be accommodated.

\begin{figure}
    \centering
    \includegraphics[width=0.5\columnwidth]{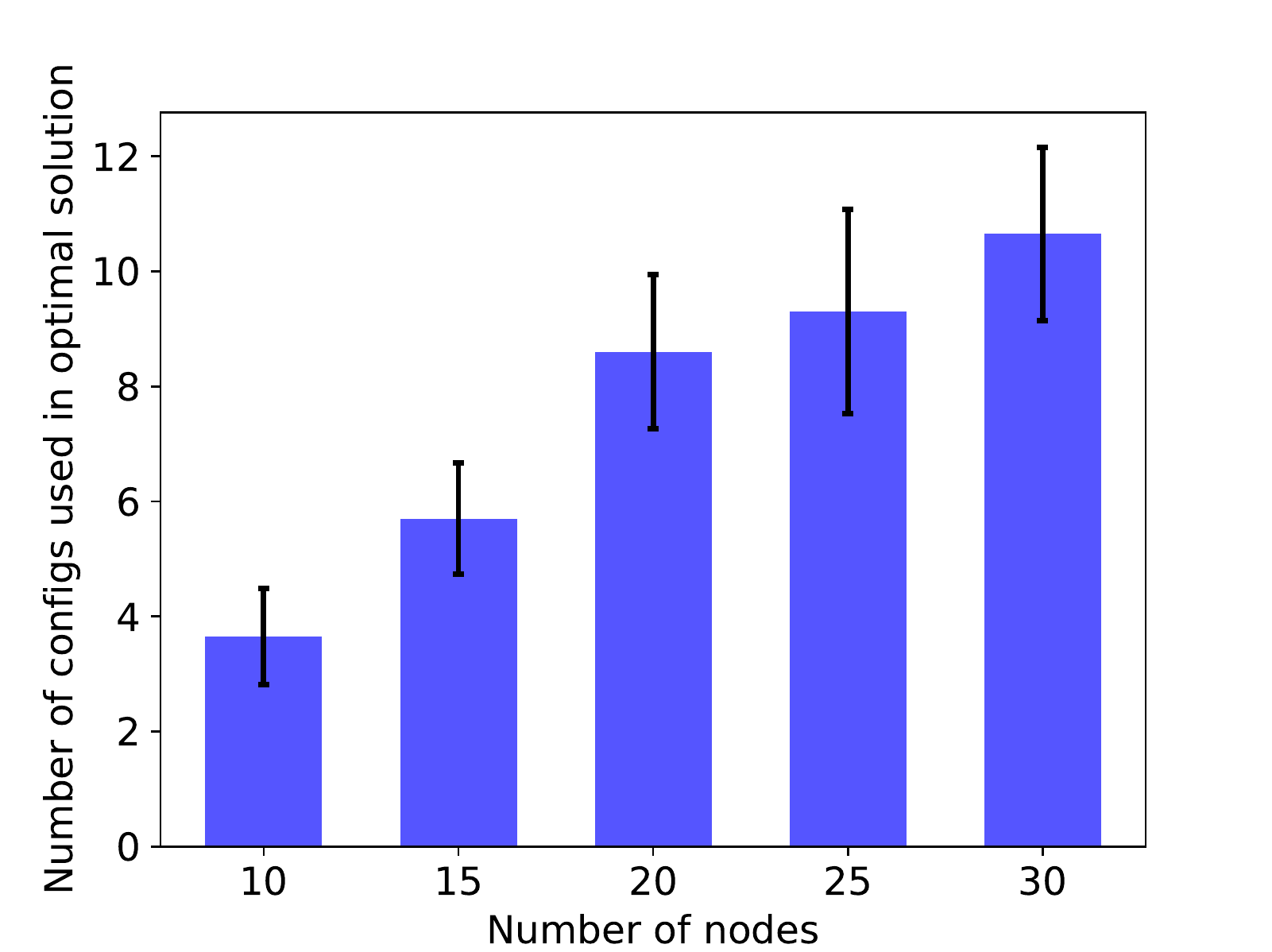}
    \caption{Number of network configurations used in the final optimal
    solution (linear relaxation).}
    \label{fig:num_configs}
\end{figure}

\begin{figure}
    \centering
    \includegraphics[width=0.5\columnwidth]{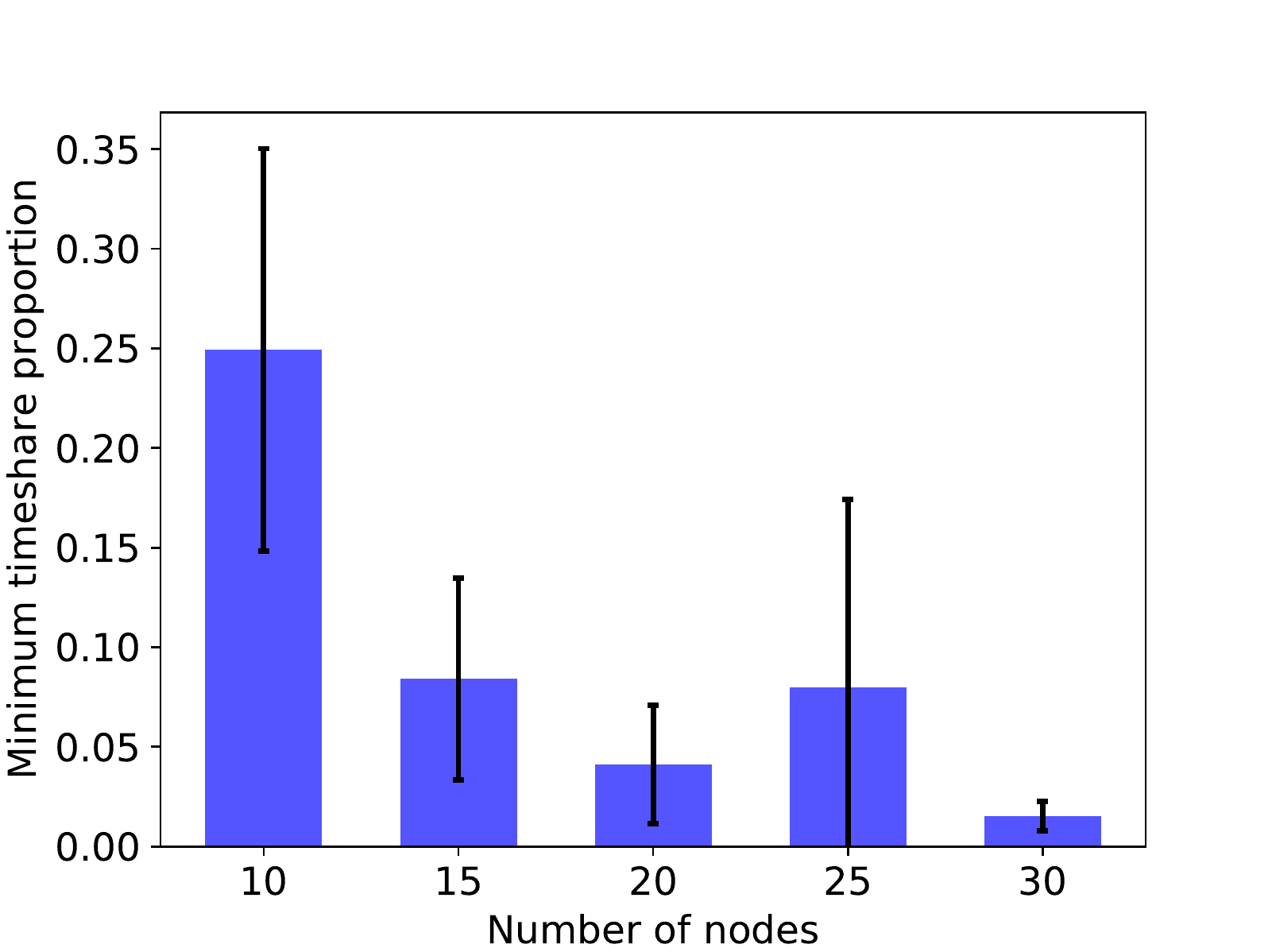}
    \caption{Minimum proportion of total network lifetime allocated to any one
    configuration used in the final optimal solution (linear relaxation).}
    \label{fig:min_timeshare}
\end{figure}

Another aspect that impacts the feasibility of implementing optimal solutions in
practice is the overhead required for switching between different network
configurations. Figure~\ref{fig:num_configs} shows the number of configurations
used in the final optimal solution as a function of network size, and
Figure~\ref{fig:min_timeshare} shows the minimum timeshare assigned to any
configuration. The number of configurations used was relatively small, around 10
even for the largest networks we tested, but did increase with the network size.
Further, as shown in Figure~\ref{fig:min_timeshare}, in some cases the
timeshares could be quite small. As such, whether or not it is worthwhile to use
these small timeshares depends on the time and energy needed to switch
configurations, as well as the absolute battery capacity, since the absolute
time (in seconds) for each timeshare scales directly with the battery capacity.
If the overall lifetime is very long, even a small timeshare --- representing
only a few percentage points of the total lifetime --- may last for days or
weeks, and therefore be worth employing in practice.

\subsubsection{Integer Problem}

\begin{figure}
    \centering
    \includegraphics[width=0.5\columnwidth]{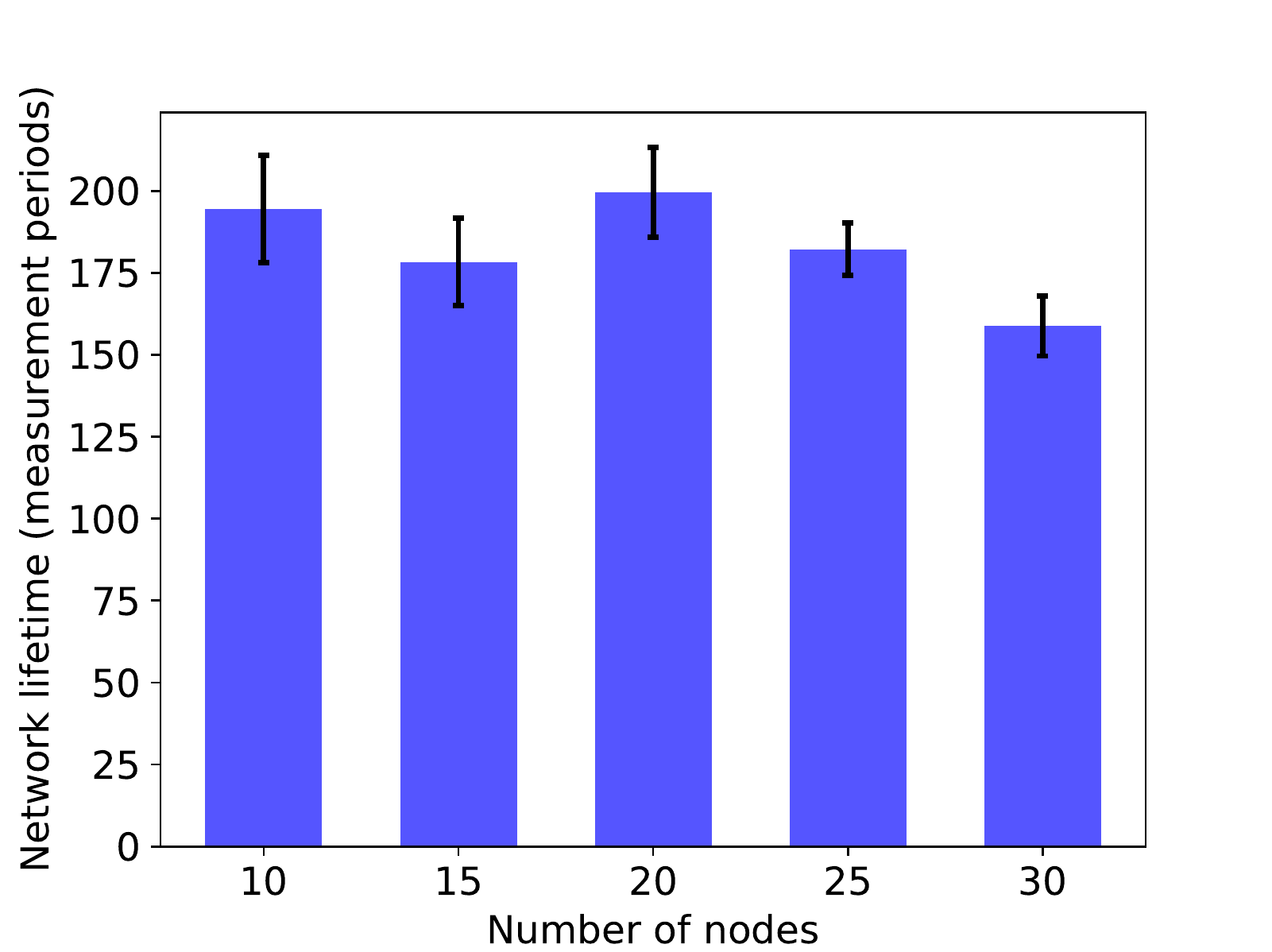}
    \caption{Network lifetime vs. number of nodes in the network (integer problem).}
    \label{fig:ip_lifetime}
\end{figure}

\begin{figure}
    \centering
    \includegraphics[width=0.5\columnwidth]{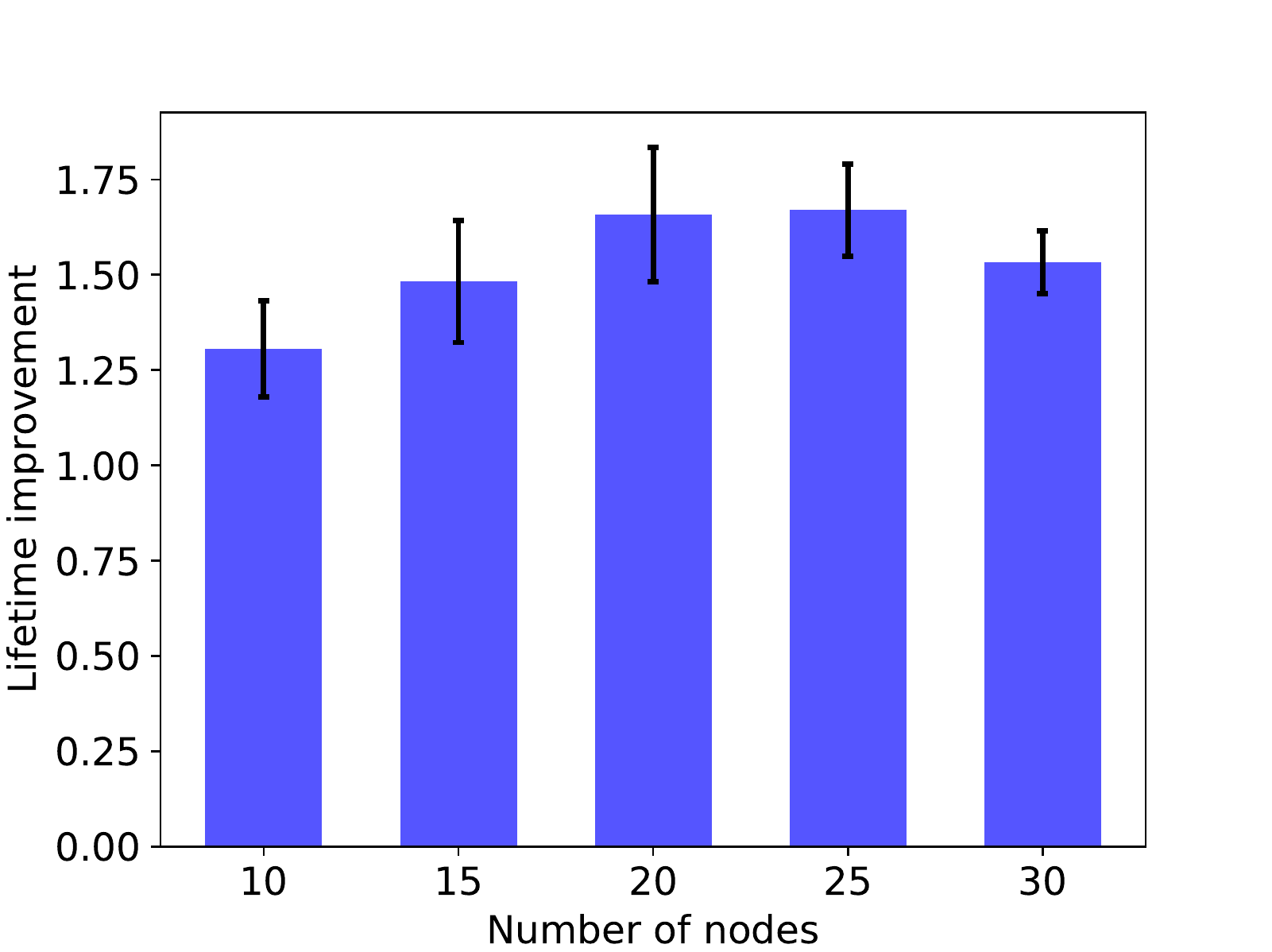}
    \caption{Network lifetime improvement compared with minimum total energy
    configuration vs. number of nodes in the network (integer problem).}
    \label{fig:ip_lifetime_improvement}
\end{figure}

In the integer case, the network lifetime (Figure~\ref{fig:ip_lifetime}) and the
lifetime improvement compared with using the minimal total energy solution
(Figure~\ref{fig:ip_lifetime_improvement}) showed similar behaviour to that seen
in the linear case. Although the highest improvement achieved was slightly lower
than for the linear relaxation, since the integer problem gives exact solutions,
we will have no further decrease in lifetime due to quantization as we do if we
apply the linear solution.

\begin{figure}
    \centering
    \includegraphics[width=0.5\columnwidth]{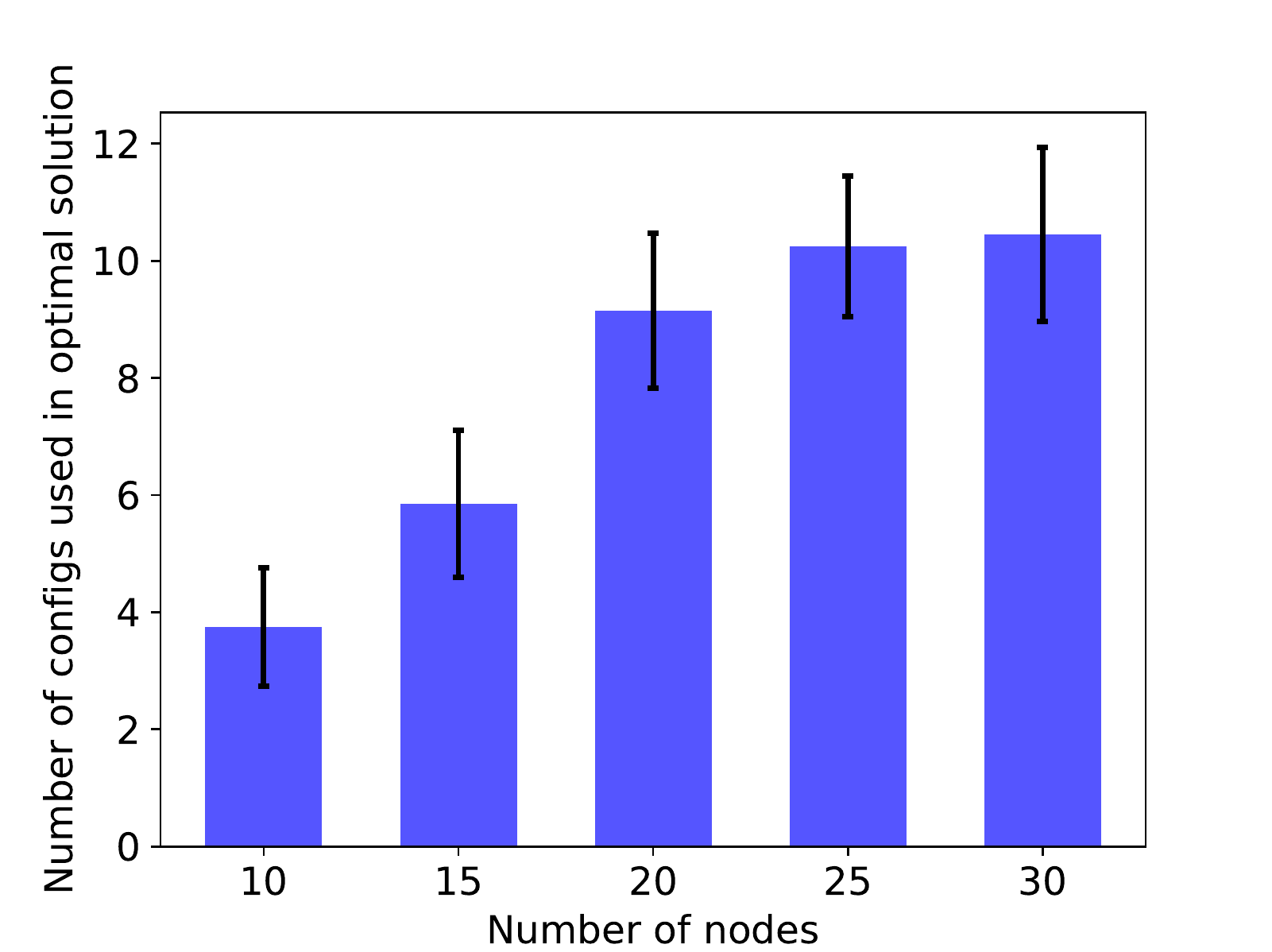}
    \caption{Number of network configurations used in the final optimal
    solution (integer problem).}
    \label{fig:ip_num_configs}
\end{figure}

\begin{figure}
    \centering
    \includegraphics[width=0.5\columnwidth]{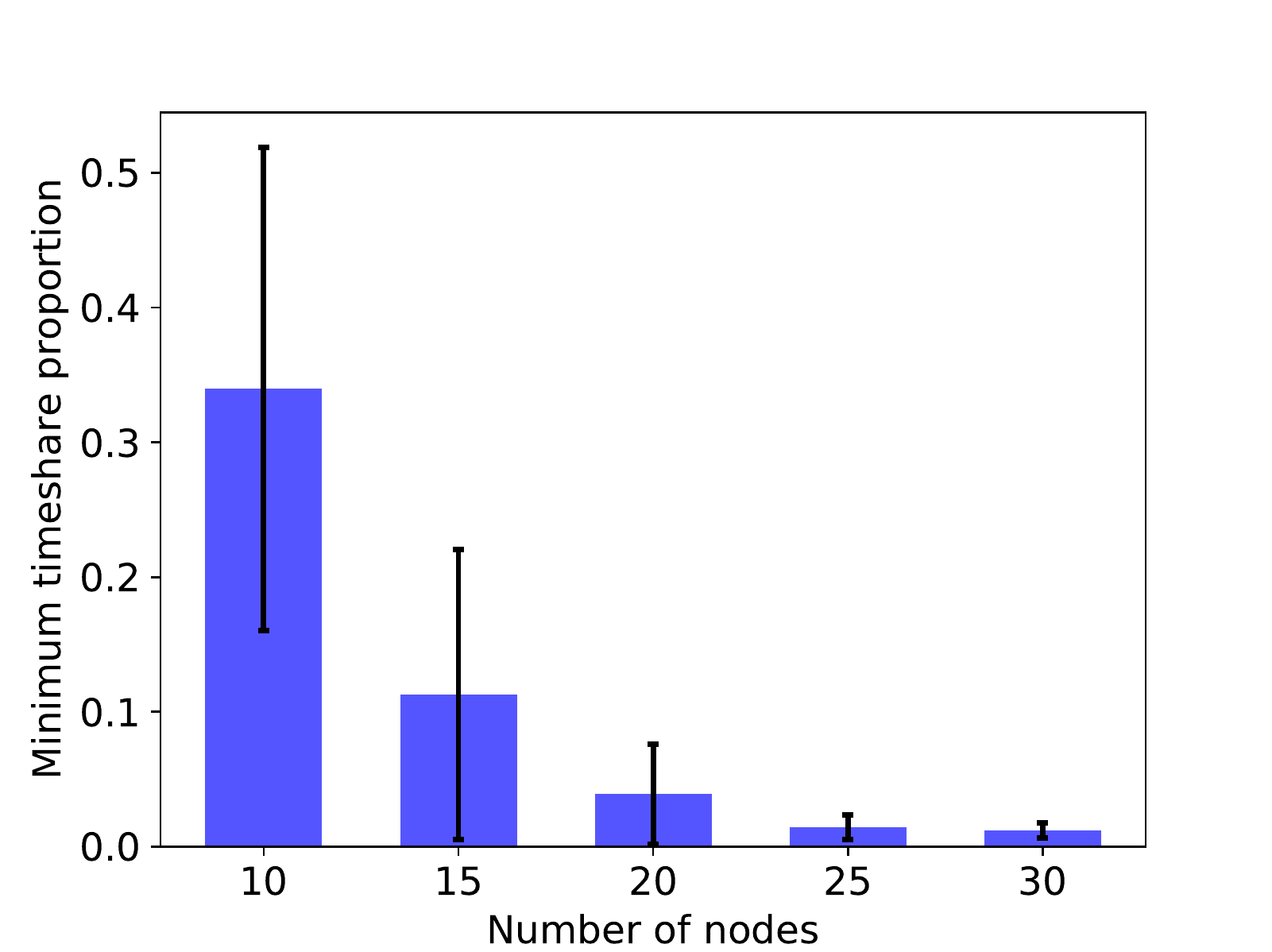}
    \caption{Minimum proportion of total network lifetime allocated to any one
    configuration used in the final optimal solution (integer problem).}
    \label{fig:ip_min_timeshare}
\end{figure}

In terms of actually solving the problem, the number of configurations actually
used in the solution to the master problem was again similar
(Figure~\ref{fig:ip_num_configs}), as were the smallest timeshares
(Figure~\ref{fig:ip_min_timeshare}). However, in the integer case, no timeshare
can be smaller than a single measurement period, so there is an inherent lower
bound on the smallest possible timeshare. Indeed, in some solutions to our test
cases the smallest timeshare was just one measurement period.

Even though the primal problem in this case used integer variables for the
timeshares, the solution times were nonetheless similar. Both the primal and
dual problems solved very quickly in all cases, with the time required dominated
by the pricing problem. (Since the column generation process is the same for
both the integer and linear cases, the solution times for the pricing problem in
the integer case must always follow the same behavior as for the linear
relaxation case.) This means that the integer version of the problem is just as
feasible to use in practice as the linear relaxation, as long as the battery
capacities of the nodes are known in advance.

\subsubsection{Comparison of solution bounds}

\begin{figure}
    \centering 
    \includegraphics[width=0.5\columnwidth]{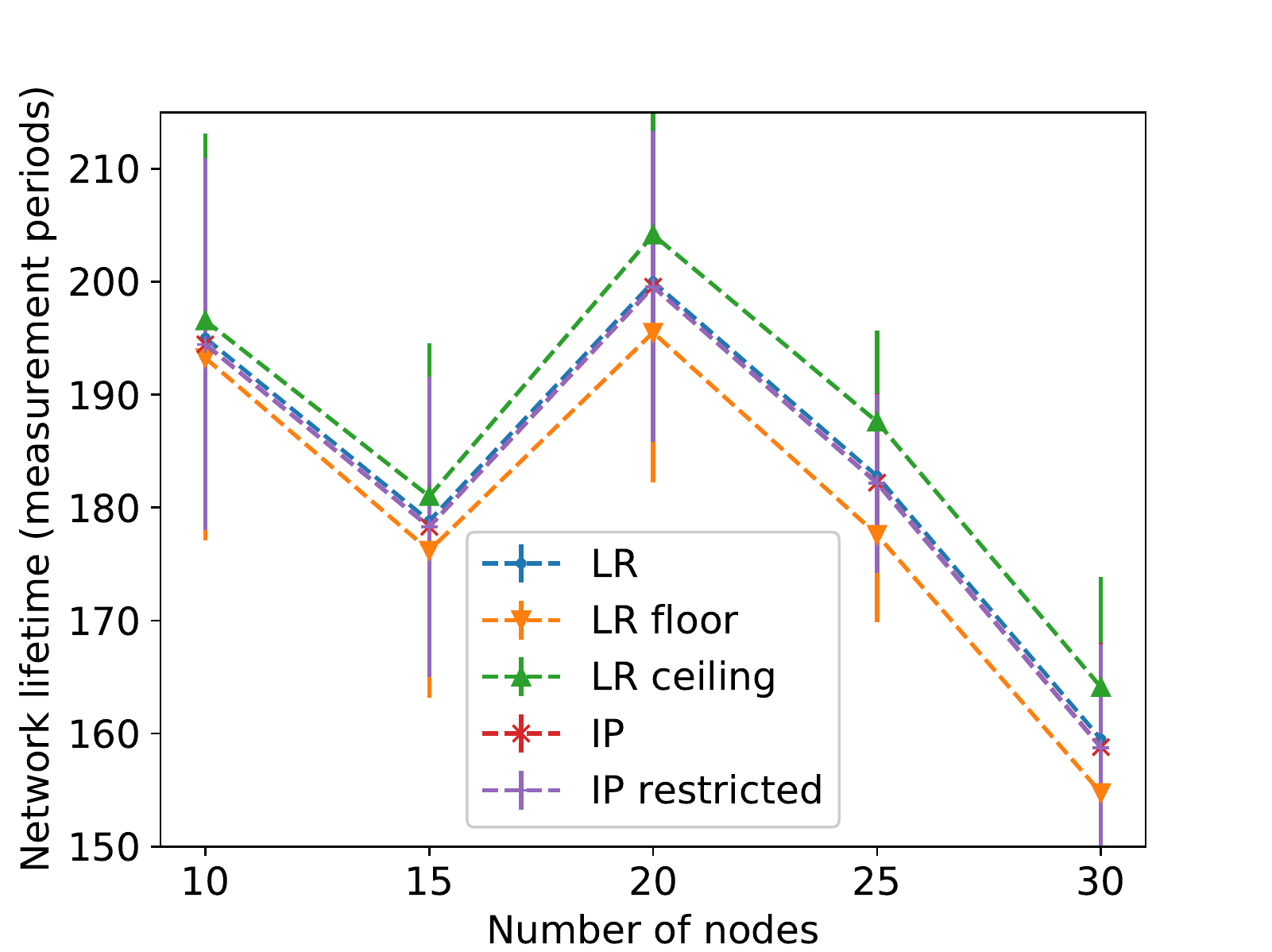}
    \caption{Network lifetime vs. number of nodes in the network, comparison of
    different linear and integer solutions.}
    \label{fig:heuristics_lifetime}
\end{figure}

As discussed in Section~\ref{sec:solve-MP}, we can readily obtain a feasible
solution to the integer version of the master problem by rounding down the
timeshares that constitute the solution to the linear relaxation. We will call
this solution the \emph{LR floor}. A better solution can be obtained by solving
the integer problem using $\mathcal C''$ --- the set of network configurations
with non-zero timeshares in the linear relaxation --- rather than all generated
configurations. We will refer to this solution as the \emph{IP restricted}
solution. Figure~\ref{fig:heuristics_lifetime} shows the network lifetime for
these two lower bounds, as well as using the linear relaxation, and an upper
bound obtained by rounding up the timeshares in the solution to the linear
relaxation, called \emph{LR ceiling}. 

As can be seen in the figure, the bounds on the integer problem are quite tight,
meaning that a good solution can be obtained using either of the proposed lower
bounds. The variation in network lifetime, as in the previous results, is
quite small, but shown in Figure~\ref{fig:heuristics_lifetime} at a larger scale
in order to distinguish the different solution methods. Since the confidence
intervals largely overlap, we cannot infer any trend in the lifetime for any
of the solution methods, save that at 30 nodes it is lower than at 10 nodes.  More data would
be needed to determine whether the lifetime would continue to decrease at larger
network sizes.

\begin{figure}
    \centering
    \includegraphics[width=0.5\columnwidth]{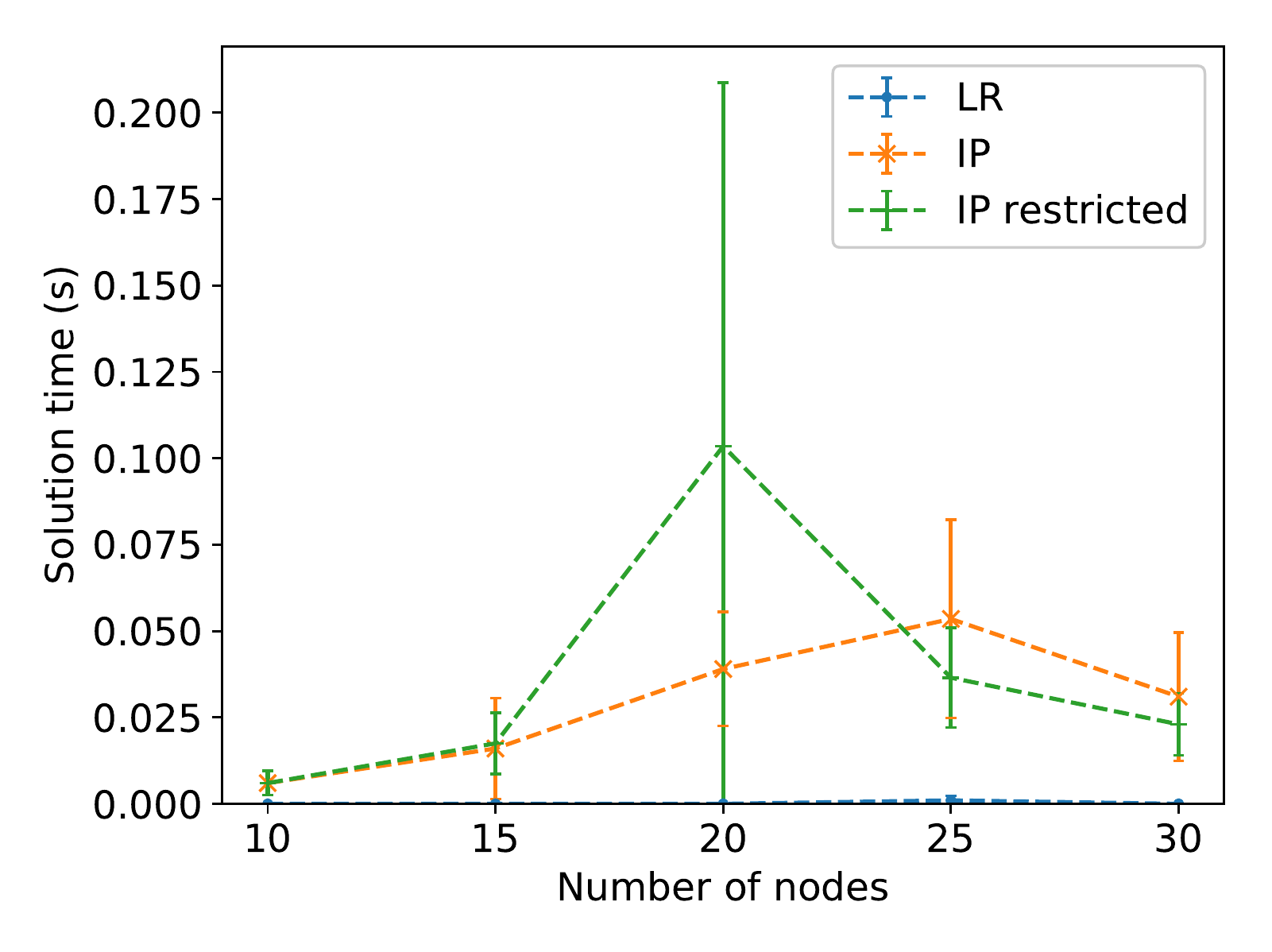}
    \caption{Solution time vs. number of nodes in the network, comparison of
    different integer solutions.}
    \label{fig:heuristics_solution_time}
\end{figure}

However, for the network sizes we tested, using the lower bounds does not result in
significant savings in solution time
(Figure~\ref{fig:heuristics_solution_time}); in fact, for 20 nodes, the average
solution time for the IP restricted solution was higher than for the IP with all
configurations. This was however due to two anomalous network instances that
took much longer to solve for the IP restricted case; for most network instances
the IP restricted solution was faster or similar to the unrestricted IP
solution. Although the solution times for all cases were very fast for our
generated networks, it may be useful to use the LR floor or IP restricted
solutions for larger networks, or for applications where the pricing problem
constitutes a smaller proportion of the overall solution time (and thus the
solution time for the master problem is more significant). 

\section{Conclusion}\label{sec:conclusion}

In this paper we have presented optimization formulations for maximizing the
network lifetime, that is, the total operating time, of a wireless mesh network.
In particular, we have focused on the case of machine-to-machine communication
requiring data aggregation and dissemination within the network, where nodes are
heterogeneous and may take on different roles and tasks. Our models allow
for reconfiguration of the network as nodes drain their batteries, thus
balancing the load of transmission and processing (aggregation) over time to
ensure the network continues to function for as long as possible.

Our numerical study, conducted on randomly generated wireless mesh networks from
10 to 30 nodes in size, shows the potential to improve the network lifetime
substantially through such intelligent reconfiguration. We achieved increases in
network lifetime of up to 75\% over the network configuration giving the minimal
total energy usage. These gains are consistent over different network sizes.
Although the number of configurations required to achieve the maximum lifetime
increased with the network size, it remained low in all cases, with only around
10 different configurations used. In applications where network lifetime is an
important performance metric, such as infrastructure or agriculture monitoring,
the network is expected to work over a longer time period of months or even
years. Reconfiguration is thus needed only infrequently in order to obtain the
maximum possible lifetime.

The models and methodology we have developed here are also general, able to
be applied to any type of network operation. We have focused on data aggregation
and dissemination, however, optimizing the network lifetime for a different task
requires only changing the pricing problem to express the constraints of the
task and its energy costs for each node. The main problem solving for the
maximum network lifetime, along with the solution approach we employ, will
continue to apply, and therefore represent useful tools for a
wide range of use cases.

\section*{Acknowledgements}

The presented work was supported by the National Science Centre, Poland, under
the grant no. 2017/25/B/ST7/02313, ``Packet routing and transmission scheduling
optimization in multi-hop wireless networks with multicast traffic''. The work
of Emma Fitzgerald was also partially supported by the Celtic-Plus project 5G
PERFECTA, the Swedish Foundation for Strategic Research project SEC4FACTORY
under the grant no. SSF RIT17-0032, and the strategic research area ELLIIT.

\balance

\section*{References}
\bibliography{references}

\end{document}